\documentclass[pra,twocolumn,showpacs,preprintnumbers,amsmath,amssymb,superscriptaddress]{revtex4-2}
\usepackage{amsmath}
\usepackage{amsfonts}
\usepackage{graphicx}
\usepackage{epsfig}
\usepackage{color}
\usepackage{textcomp}
\usepackage{physics}
\usepackage{wasysym}
\usepackage{float}
\usepackage{blindtext}
\usepackage{mathtools}
\usepackage[colorlinks,citecolor=blue]{hyperref}

\begin{document}
	
	\title{Power-Law-Exponential Interaction Induced Quantum Spiral Phases}
	\author{Guoqing Tian}
	\affiliation{School of Physics and Institute for Quantum Science and Engineering,
Huazhong University of Science and Technology and Wuhan Institute of Quantum Technology, Wuhan 430074, China}

	\author{Ying Wu}
	\affiliation{School of Physics and Institute for Quantum Science and Engineering,
Huazhong University of Science and Technology and Wuhan Institute of Quantum Technology, Wuhan 430074, China}
	
	\author{Xin-You L\"{u}}\email{xinyoulu@hust.edu.cn}
	\affiliation{School of Physics and Institute for Quantum Science and Engineering,
Huazhong University of Science and Technology and Wuhan Institute of Quantum Technology, Wuhan 430074, China}

\date{\today}
	
\begin{abstract}
We theoretically predict a kind of power-law-exponential (PLE) dipole-dipole interaction between quantum emitters in a 1D waveguide QED system. This unconventional long-range interaction is the combination of power-law growth and exponential decay couplings. Applying PLE interaction to a spin model, we uncover the rich many-body phases. Most remarkably, we find that PLE interaction can induce the ordered and critical spiral phases. These spiral phases emerge from the strong frustration generated by the power-law factor of PLE interaction, hence they are absent for other types of long-range interaction, e.g., pure exponential and power-law decay interactions. Our work is also applicable for the higher dimensional systems. It fundamentally broadens the realm of many-body physics and has the significant applications in quantum simulation of strong correlated matters.
	\end{abstract}
	\maketitle

\section{introduction}
The studies on many-body behaviors of quantum matters are fundamentally important both in theoretical and experimental respects. In the framework of Ginzburg-Landau theory, the many-body interactions play a key role to induce intriguing phenomena, i.e., quantum phases and phase transitions\,\cite{giamarchi2003quantum,sachdev2011quantum,auerbach}. Significant progress has been made in understanding many-body phases in systems with short-range interactions, while long-range interacting systems have continuously attracted research attention as they can induce distinctive phases without counterparts in short-range interacting systems\,\cite{dauxois2002dynamics,campa2014physics,CAMPA200957,lahaye2009physics,saffman2010quantum,RevModPhys.95.035002}. For example, long-range interactions can give rise to Wigner crystallization\,\cite{wigner}, and continuous symmetry broken in low-dimensional systems\,\cite{Mermin,Vodola2015,Maghrebi}.

Waveguide QED system where QEs couple to a structured bosonic environment, provide a platform to efficiently generate long-range interactions\,\cite{John1990,douglas2015quantum,AJH1,Giuseppe2016,manzoni2017designing,AJH2}. In particular, mediated by a 1D bipartite photonic lattice, the spatial profile of emergent long-range dipole-dipole (d-d) interactions between QEs exhibits the topological edge state features as the frequency of QEs is resonant with the edge state energy of lattice\,\cite{Bello2019,KimTopological,Leonforte2021,Cheng}. Especially, the non-monotonic d-d interaction emerges in the extended SSH lattice\,\cite{Vega}. This provides an opportunity to explore exotic phases of matter induced by these long-range interactions\,\cite{Bellophase}.
	
In this work, we investigate the system consists of QEs coupling to a 1D bipartite photonic lattice endowed with chiral symmetry. By integrating hopping terms beyond nearest-neighbor into lattice and appropriately designing the hopping strengths, we predict a kind of power-law-exponential d-d interactions featuring with non-monotonicity. Different from the previous work\,\cite{Vega} reporting the d-d interaction featuring non-monotonic spatial behavior, here the non-monotonicity of PLE interactions entirely comes from the power-law increasing term $x^{\alpha}$. Mathematically, such power-law factor originates from the existence of higher order zeros in the characteristic polynomial of bath. Physically, it is the consequence of the interference between multiple exponential decay components of interaction with the same decay length.
	
More importantly, applying the long-range PLE interaction to an effective spin model, we uncover the rich many-body phases in this system for the first time. Especially, as the interaction length increasing, we find a quantum critical phase, i.e., the quasi-long-range ordered phase (QLRO(T)) with T indicating the wave number of the dominant spin correlations. Distinctive from the other critical phase occurring for short interaction length, the spin-spin correlation function in QLRO(T) phase is spiral with distance, and has the incommensurate period T. In addition, an ordered spiral phase named antiferromagnetic phase with period T (AFM(T)) also appears when the spin-exchange interaction is ignorable in comparison with $zz$ interaction. Similar to the QLRO(T) phase, the spin arranges antiferromagnetically in this AFM(T) phase but with commensurate periods. We explain that these spiral phases emerge from the strong frustration induced by the power-law increasing factor of interaction, and thus has no analog in other kinds of long-range interaction (e.g., exponential and power-law decay interactions) systems. Our work open up a door for exploring unconventional long range interactions by employing the topological property of bath, which is fundamentally significant in realizing the novel many-body phases.

The paper is structured as follows. In Sec.\,\ref{II}, we theoretically predict the emergence of PLE interaction in generic photonic waveguide endowed with chiral symmetry, explain the associating physical mechanism by a concrete model and discuss the robustness of PLE interaction against disorder. In Sec.\,\ref{III}, we investigate the spin many-body phase diagram for system with PLE interaction, and compare it with the phase diagram for systems with exponential and power-law decay interactions. In Sec.\,\ref{IV}, we discuss the experimental implement of PLE interaction based on the recently experimental and theoretical proposals. Finally, we summarize our findings in Sec.\,\ref{V}.

\section{PLE d-d interaction mediated by chirally symmetric lattice}\label{II}
\subsection{Unified form of d-d interaction in the VDS mechanism}
\begin{figure}
		\centering
		\includegraphics[width=8.4cm]{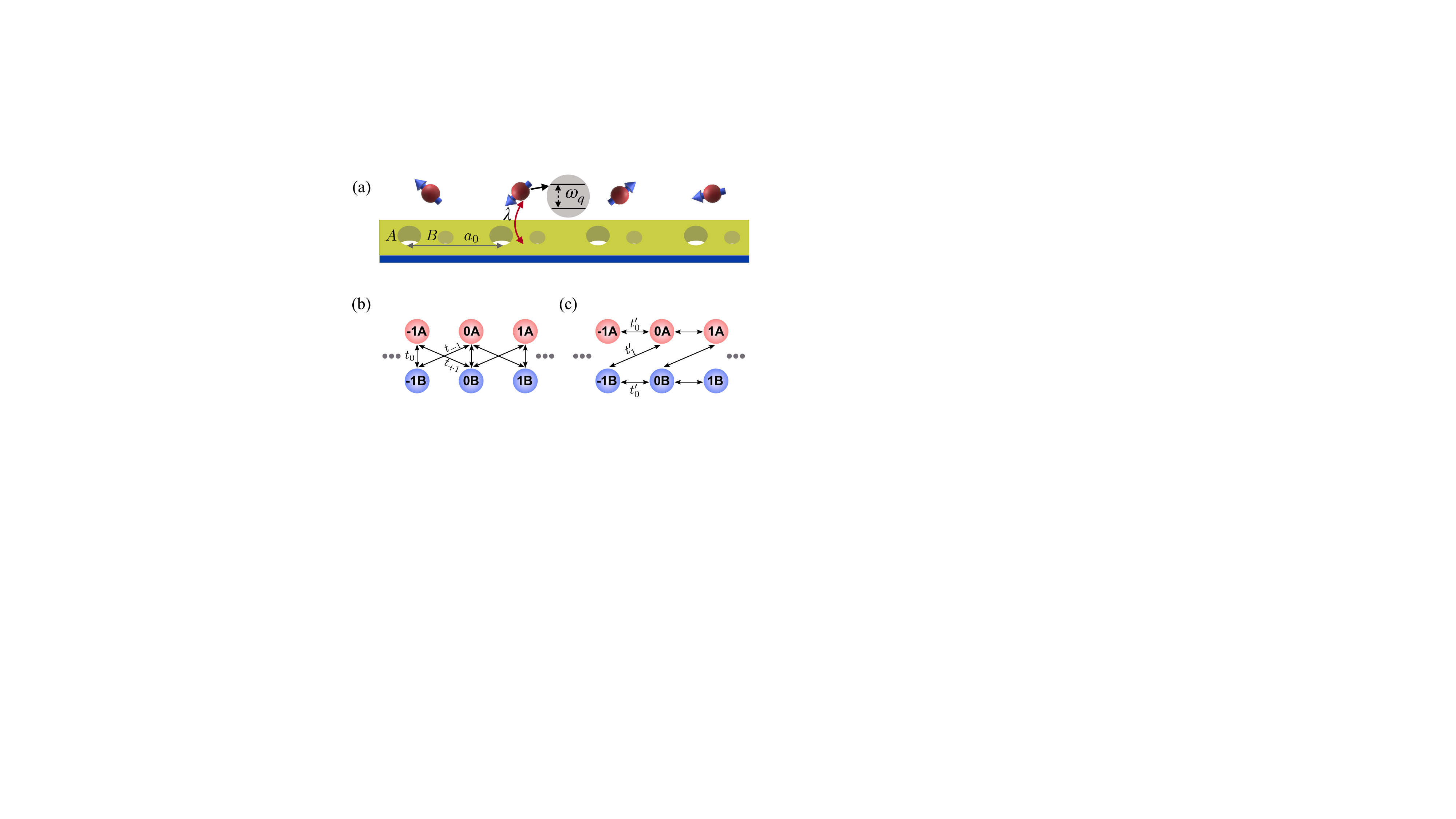}
		\caption{(a) $N$ two-level QEs coupled to 1D coupled-cavity array, where two adjacent cells are separated by $a_0$. (b) Real space configuration of the lattice bath for $h(k)=t_0+t_{-1}\exp(-ik)+t_{+1}\exp(ik)$ with the hopping strengths $\qty{t_0;t_{-1};t_{+1}}$ and thus $P=1,\ Q=1$. (c) Real space configuration of the lattice bath for $H(k)=t'_1\sin(k)\sigma_y+(t'_0\cos(k)+m_0)\sigma_z$, which respects the chiral symmetry $\sigma_x$ and is able to produce the d-d interaction without chirality.}\label{fig1}
	\end{figure}

As shown in Fig.\,\ref{fig1}(a), we consider $N$ two-level QEs coupled to 1D photonic two-sublattice bath. The total Hamiltonian of system reads (setting $a_0=1$)
\begin{align}\label{eq1}
	\!\!\!\!H_{N}=\omega_q\sum_{j=1}^{N}{\sigma^{\dagger}_j\sigma_j}\!+\!H_b\!+\!\lambda\sum_{j=1}^{N}(a^{\dagger}_{x_{j,\alpha}}\sigma_j+h.c.),\!\!
\end{align}
where $\omega_q$ is the transition frequency of QEs, and $\lambda$ is the coupling strength. Moreover, $\sigma_j=|g\rangle_j\langle e|$ is the Pauli annihilation operator for the $j$-th emitter, and $a_{x_{j,\alpha}}$ is the annihilation operator corresponding to the position $x_{j,\alpha}$ of bath (i.e., sublattice $\alpha\in\left\{A,B\right\}$ at the $x_{j}$-th cell). Under periodic boundary condition, the bath Hamiltonian is given by $H_b\!=\!\sum_k\mqty(a^{\dagger}_{k,A},\!&\!a^{\dagger}_{k,B})H_b(k)\mqty(a_{k,A},\!&\!a_{k,B})^T$ with $H_b(k)=\omega_0\mathbb{I}+d_x(k)\sigma_x+d_y(k)\sigma_y+d_z(k)\sigma_z$. We consider the bath respecting a chiral symmetry $\sigma_z(H_b(k)-\omega_0\mathbb{I})\sigma_z=-(H_b(k)-\omega_0\mathbb{I})$ (hence $d_z(k)\!=\!0$). The full property of the lattice bath is encoded in $h(k)=d_x(k)-id_y(k)$, whose generic form can be written as $h(k)=\sum_{m=0}^{P}t_{-m}e^{-imk}+\sum_{n=0}^{Q}t_{+n}e^{ink}$ with the hopping strengths of lattice $\{t_0;t_{-1},\cdots,t_{-P};t_{+1},\cdots,t_{+Q}\}$. The integer $P$ ($Q$) denotes the at most $P$-site ($Q$-site) hopping from sublattice A to B towards the left (right) direction [see Fig.\,\ref{fig1}(b) for an example].
	
	In the single-excitation regime, when a single QE with frequency $\omega_q=\omega_0$ couples to bath at $x_{1,\alpha}$, a vacancy-like bound state with eigenenergy $E=\omega_0$ occurs\,\cite{Leonforte2021}. When the second QE located at $x_{2,\beta}$ is integrated into the bath, the mechanism of vacancy-like bound state guarantees the resulted d-d interaction has the same spatial profile with the zero-energy state of $H_b$ (we set $\omega_q=\omega_0=0$ for simplify), i.e., $J(x_{1,\alpha},x_{2,\beta})\propto\lambda\braket{x_{2,\beta}}{\varphi_E}$, where $\ket{\varphi_E}$ is the zero-energy solution of bath under hard-core boundary condition\,\cite{Leonforteimpurities}. Without loss of generality, we assume that the first QE couples to bath at sublattice A of the 0-th cell. We ignore the case of which the second QE couples to bath at sublattice A, since the obtained d-d interaction is zero due to the chiral symmetry $\sigma_z$, and we use the short notation $J(x)=J(x_{2,B}-x_{1,A})\equiv J(x_{1,A},x_{2,B})$. In thermodynamic limit of bath, we obtain a unified form of d-d interaction by solving the zero-energy state. The expression is given by Appendix\,\ref{suppA}
	\begin{equation}\label{int}
		\frac{J(x)}{\lambda}\propto
		\left\{
		\begin{aligned}
			&\sum_{\mu=1}\sum_{\alpha=0}^{n_{\mu}-1}r_{\mu\alpha}(x-x_c)^{\alpha}e^{-(x-x_c)/\xi_{\mu}},x\ge x_c, \\
			&\sum_{\nu=1}\sum_{\beta=0}^{n_{\nu}-1}l_{\nu\beta}(x-x_c)^{\beta}e^{-(x-x_c)/\xi_{\nu}},x<x_c,
		\end{aligned}
		\right.
	\end{equation}
where $x_c=1-P$, $r_{\mu\alpha}$ and $l_{\nu\beta}$ are constants, and the summation indexes $\mu,\nu$ cover all distinctive zeros $z_{\mu,\nu}$ of the characteristic polynomial $h(z)$ obtained from $h(k\!\to\!-i\ln z)$ in complex plane. Here $n_{\mu,\nu}$ is the order of $z_{\mu,\nu}$, and $\xi_{\mu,\nu}=-\ln^{-1}(z_{\mu,\nu})$ is the decay length of interaction. All physical quantities mentioned here are exactly solvable and are determined by the bulk property of bath.

Since the power-law factor $(x-x_c)^{\alpha,\beta}$ in Eq.\,(\ref{int}), the strength of PLE interaction is weakened at short range and enhanced at long range comparing with the pure exponential decay interaction. Its maximal strength locates at $x=\alpha\xi_{\mu}-x_c$.  The occurrence of power-law factor originates mathematically from the higher-order zero of $h(z)$. Since the maximal order of zero should be less than or equal to the number of zeros, we obtain the upper bounds on the power-law exponents
	\begin{align}\label{bound}
		\max_{\Re\xi_{\mu}>0}\qty{n_{\mu}}\leq P+W,\ \max_{\Re\xi_{\nu}<0}\qty{n_{\nu}}\leq Q-W,
	\end{align}
where $W=(2\pi i)^{-1}\int_0^{2\pi}\partial_k\log(h(k))dk$ is the winding number characterizing the topology of bath. Physically, the value of $P+W$ ($Q-W$) counts the number of zeros of $h(z)$ lying within (outside) the unit circle, and then it is equal to the number of exponential decay components involved in the superposition of d-d interaction. When the lattice is further engineered such that $n_{\mu}$ ($n_{\nu}$) out of $P+W$ ($Q-W$) exponential decay components of interaction possess the same decay length, the interference between $n_{\mu}$ ($n_{\nu}$) exponential decay components leads to the PLE interaction towards to $x>x_c$ ($x<x_c$) direction with maximal exponent $n_{\mu}-1$ ($n_{\nu}-1$). The detailed calculation and discussion are shown in Supplemental Material Appendix\,\ref{suppA}. 
	
\subsection{Minimal model exhibiting PLE interaction}
We present an example to demonstrate the physical mechanism of realizing PLE interaction more clearly. To this end, considering an extended SSH bath with $h(k)=t_0+t_{-1}\exp(-ik)+t_{+1}\exp(ik)$. In addition to the nearest neighbor hopping $t_0$ and $t_{-1}$, we introduce the next-next (NN) nearest hopping $t_{+1}$ into bath. For hopping strengths $t^2_0\ne4t_{+1}t_{-1}$, the d-d interaction mediated by this bath yields Appendix.\,\ref{suppA}
	\begin{equation}\label{2roots}
		\frac{J(x)}{\lambda}\propto
		\left\{
		\begin{aligned}
			&\frac{r(\xi_1)e^{-x/\xi_1}-r(\xi_2)e^{-x/\xi_2}}{e^{-1/\xi_1}-e^{-1/\xi_2}},x\ge 0 \\
			&\frac{l(\xi_1)e^{-x/\xi_1}-l(\xi_2)e^{-x/\xi_2}}{e^{-1/\xi_1}-e^{-1/\xi_2}},x<0
		\end{aligned}
		\right.
	\end{equation}
where $r(z)=\!\Theta(1\!-\!|\exp(1/z)|){\rm sign}(1\!-\!|\exp(1/z)|)$, $l(z)\!=\!\Theta(-1+|\exp(1/z)|){\rm sign}(1-|\exp(1/z)|)$, and $\xi_{1,2}\!=-\!1/\ln(z_{1,2})$. Here $z_{1}=(-t_0+(t^2_0-4t_{+1}t_{-1})^{1/2})/2t_{+1}$, $z_{2}=(-t_0-(t^2_0-4t_{+1}t_{-1})^{1/2})/2t_{+1}$ are the zeros of $h(z)$, and $\Theta(x)$ is the Heaviside function. It is shown from Eq.\,(\ref{2roots}) that a zero lying within (outside) the unit circle contributes an exponential decay component. Thus the d-d interaction is the superposition of two exponential decay components, when two zeros lie simultaneously within or outside the unit circle. As the hopping strengths being engineered such that the condition for PLE interaction $4t_{+1}t_{-1}=t^2_0$ is satisfied, two zeros coincide and Eq.\,(\ref{2roots}) becomes intermediate. Then, one needs to take the limit to obtain the interaction (suppose $\Re(\xi_{1,2})>0$), i.e., 
	\begin{align}\label{ple}
		\frac{J(x\!\ge\!0)}{\lambda}\propto\lim_{\xi_{1,2} \to \xi}\frac{e^{-x/\xi_1}-e^{-x/\xi_2}}{e^{-1/\xi_1}-e^{-1/\xi_2}}\propto xe^{-x/\xi}.
	\end{align}
	This formula indicates the underlying interference mechanism for PLE interaction with power-law exponent $1$, namely the out-of-phase superposition of two exponential decay components with the same decay length $\xi$.
	
	\begin{figure}
		\centering
		\includegraphics[width=8.4cm]{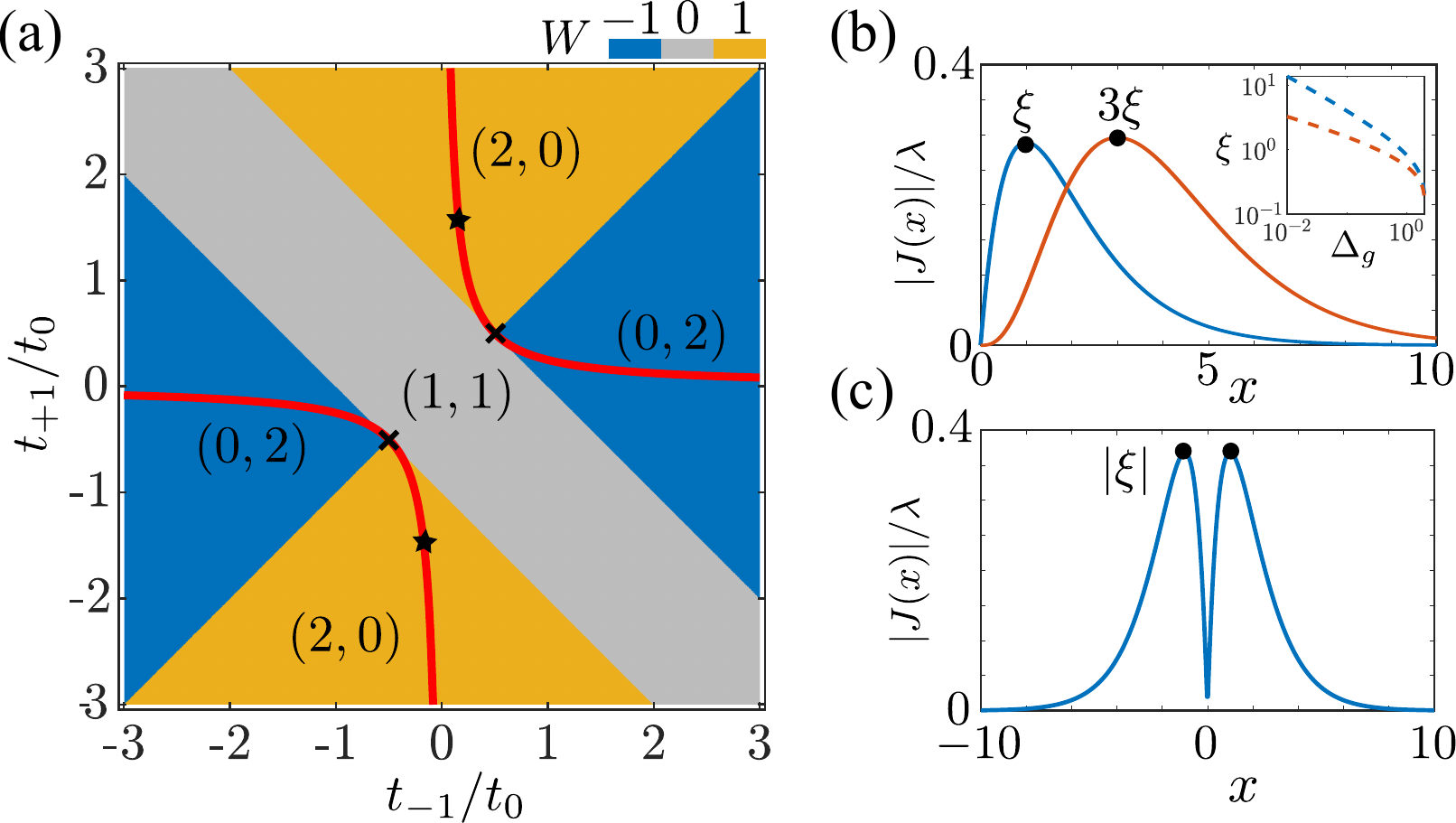}
		\caption{(a) Phase diagram of the lattice bath. Red line $4t_{+1}t_{-1}=t^2_0$ is the parameter regime for PLE interaction. The cross marks label two tri-phase points, where the PLE interaction is absent. The integer pairs are the upper bounds $(P+W,Q-W)$. (b) Spatial profile of PLE interaction $x\exp(-x/\xi)$ (blue line) with right-hand chirality that corresponds to the parameters denoted by black stars in (a). The orange line displays PLE interaction $x^3\exp(-x/\xi)$, the lattice configuration is given in Appendix\,\ref{suppB}. Inset displays the decay length $\xi$ versus the band gap of lattice bath $\Delta_g/t_0$ for power-law factor $1$ (dashed blue line) and $3$ (dashed orange line). (c) PLE interaction without chirality $\abs{x}\exp(-\abs{x}/\xi)$ with lattice configuration giving in Fig.\,\ref{fig1}(c). In (b-c), the dots mark the positions of maximum interaction strength.}\label{fig2}
	\end{figure}
	
The phase diagram in Fig.\,\ref{fig2}(a) illustrates the parameter regimes for PLE interaction. Owing to the additional NN nearest hopping, the lattice bath has three phases distinguished by winding number $0$ (topologically trivial) and $\pm1$ (topologically non-trivial). As indicated by Eqs.\,(\ref{int}-\ref{bound}), the upper bounds larger than $1$ is the necessary condition for the implement of PLE interaction. Therefore, the parameter regimes for PLE interaction only appear in the topological phases in which one of the upper bounds is large than 1. Within these regimes, the hopping strengths should be properly engineered and satisfy the red lines to obtain PLE interaction. Fig.\,\ref{fig2}(b) shows the spatial profile of PLE interactions with power-law factors $1$ and $3$. As an indicator of non-monotonicity, the decay length $\xi$ is controlled by the band gap of lattice according to the functional relation $\xi=-1/\ln(1-(\Delta_g/2)^{1/\nu})$, where $\nu=n_{\mu}+1$ for lattice producing power-law factor $n_{\mu}$\,\cite{def}. Note that, PLE interaction without chirality can also be obtained by coupling QEs to lattice bath with chiral symmetry $\sigma_x$ [see Fig.\,\ref{fig2}(c)] Appendix\,\ref{suppA}. Besides, our result in Supplemental Material Appendix\,\ref{suppC} show the realization of similar PLE interaction in higher dimensional lattice.
	
Based on this explicit model, the necessity of long-range hopping beyond nearest neighbor for the realization of PLE interaction becomes clear. Integrating longer-rang hopping (increasing $P$ and $Q$) into lattice will increase the values of upper bounds when keeping $W$ invariant, and thus increase the number of components involving superposition. In other words, one needs to integrate longer range hopping into the lattice to produce potential PLE interaction with higher power-law exponents Appendix\,\ref{suppB}. While for a SSH lattice (i.e., $t_{+1}=0$), the upper bounds are at most equals to $1$, which eventually leads to a pure exponential decay d-d interaction\,\cite{Bello2019}.

\subsection{Robustness against disorder}
\begin{figure*}
	\centering
	\includegraphics[width=16cm]{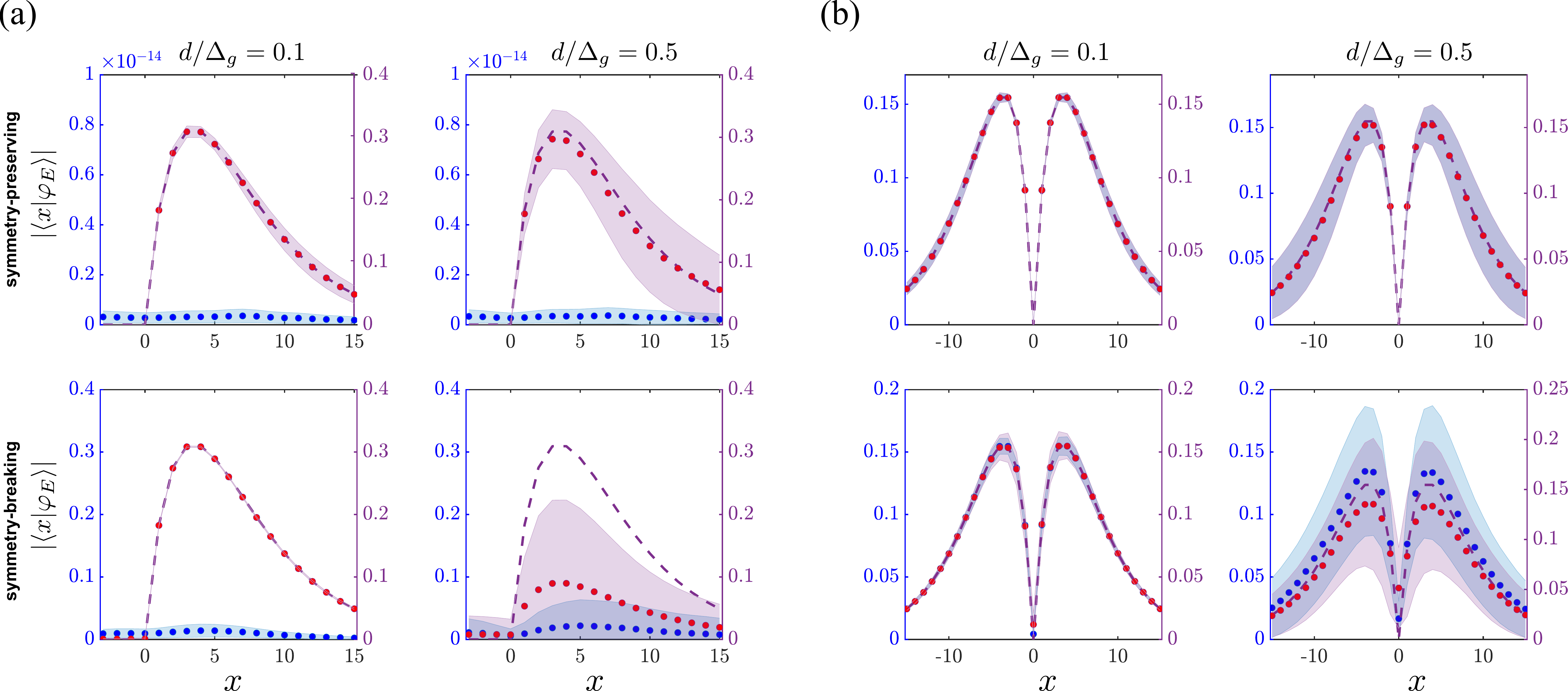}
	\caption{(a-b) PLE dressed bound states for bath Hamiltonian respecting chiral symmetry $\sigma_z$ (a) and chiral symmetry $\sigma_x$, respectively. The red (blue) dots denote components on subplattice B (A) under $10^3$ disorder realizations, while shadow areas span their corresponding standard deviation. Dashed lines are PLE state in the clean limit, and we only plot clean PLE state on sublattice B in (a) since the zero components on sublattice A. In all plots, parameters are chosen to be $\lambda/\Delta_g=2.3$, and $\Delta_g=0.13$. }\label{disorder}
\end{figure*}

In this section, we explore whether the PLE interaction inherits the topological protection feature from edge state. To do that, we study the robustness of the topological dressed bound state against disorder since this dressed bound state possesses the same spatial profile with PLE interaction. We consider two types of disorder depending on whether they preserve the chiral symmetry or not and present the results in Fig.\,\ref{disorder}. As expected, the topological bound states exhibit robustness against symmetry-preserving disorder. Besides, the shape of dressed bound states is more sensitive to symmetry-breaking disorder in comparison with symmetry-preserving disorder. Nonetheless, the dressed bound states retain qualitatively the non-monotonic behavior for both symmetry-preserving and symmetry-breaking disorder.

Specifically, for bath Hamiltonian given in Fig.\,\ref{fig1}(b) of main text, the chiral symmetry is $\sigma_z$, the obtained dressed bound states exhibit unidirectional spatial profile and has support in one of the sublattices in the clean limit. The disorder is implemented as $H_b\to H_b+\sum_j\varepsilon_{1,j}a^{\dagger}_{j,B}a_{j+1,A}+\sum_j\varepsilon_{2,j}a^{\dagger}_{j,A}a_{j,B}+\sum_j\varepsilon_{3,j}a^{\dagger}_{j,A}a_{j+1,B}+h.c$ for symmetry-preserving disorder and $H_b\to H_b+(\sum_j\varepsilon_{1,j}a^{\dagger}_{j,A}a_{j,A}+\sum_j\varepsilon_{2,j}a^{\dagger}_{j,B}a_{j,B})/2+\sum_j\varepsilon_{3,j}a^{\dagger}_{j,A}a_{j+1,A}+\sum_j\varepsilon_{4,j}a^{\dagger}_{j,B}a_{j+1,B}+h.c$ for symmetry-breaking disorder.
The coefficients $\varepsilon_{1-4,j}$ are i.i.d random variables, drawing from Gaussian distribution $\mathcal{N}(0,d)$. For symmetry-preserving disorder, the topological properties of dressed bound state, i.e the unidirectionality and the zero component on sublattice A [see the top of Fig.\,\ref{disorder}(a)], retain as expected from the mechanism of vacancy-like dressed bound state. For symmetry-breaking disorder, the resulted dressed bound state lost its unidirectionality and has non-zero component on each sublattices [see the bottom of Fig.\,\ref{disorder}(a)]. Moreover, the shape of bound state retains its non-monotonic behavior both for symmetry-preserving and symmetry-breaking disorder. The shape is more sensitive to symmetry-breaking disorder comparing with symmetry-preserving disorder, i.e., it has larger deviation from its clean counterpart for symmetry-breaking disorder.

For bath Hamiltonian given in Fig.\,\ref{fig1}(b) of main text, the chiral symmetry is $\sigma_x$. As a consequence of chiral symmetry $\sigma_x$, the obtained dressed bound state has equal weight on each sublattice of lattice. The disorder is implemented as $H_b\to H_b+\sum_j\varepsilon_{1,j}a^{\dagger}_{j,A}a_{j,A}-\sum_j\varepsilon_{1,j}a^{\dagger}_{j,B}a_{j,B}$ for symmetry-preserving disorder and $H_b\to H_b+(\sum_j\varepsilon_{1,j}a^{\dagger}_{j,A}a_{j,A}+\sum_j\varepsilon_{2,j}a^{\dagger}_{j,B}a_{j,B})/2+\sum_j\varepsilon_{3,j}a^{\dagger}_{j,A}a_{j,B}+h.c$ for symmetry-breaking disorder. The coefficients $\varepsilon_{1-3,j}$ are i.i.d random variables, drawing from Gaussian distribution $\mathcal{N}(0,d)$. For symmetry-preserving disorder, the dressed bound states have equal components on sublattice A and B at each unit cell [see the top of Fig.\,\ref{disorder}(b)]. When the introduced disorder breaks the chiral symmetry of bath Hamiltonian, the dressed bound states lost the property [see the bottom of Fig.\,\ref{disorder}(b)]. Similar to the former case, the shape of bound state retains its non-monotonic behavior both for symmetry-preserving and symmetry-breaking disorder.

\section{Spin many-body phases induced by PLE interaction}\label{III}
The unique feature of PLE interaction in comparison with other types of long-range interaction, namely the power-law growth factor, inspires us to explore the emergent many-body phases. We consider a 1D long-range XXZ model described by the Hamiltonian $H_{XXZ}=\sum^{N}_{m>n}(m-n)\exp(-(m-n)/\xi)(S^{+}_{m}S^-_{n}/2+S^{+}_{n}S^-_{m}/2+J_zS^z_mS^z_n)$. The required $zz$ interaction can be realized by the fast single-qubit rotations\,\cite{zzint1,zzint2}.

	\begin{figure}
		\centering
		\includegraphics[width=8.4cm]{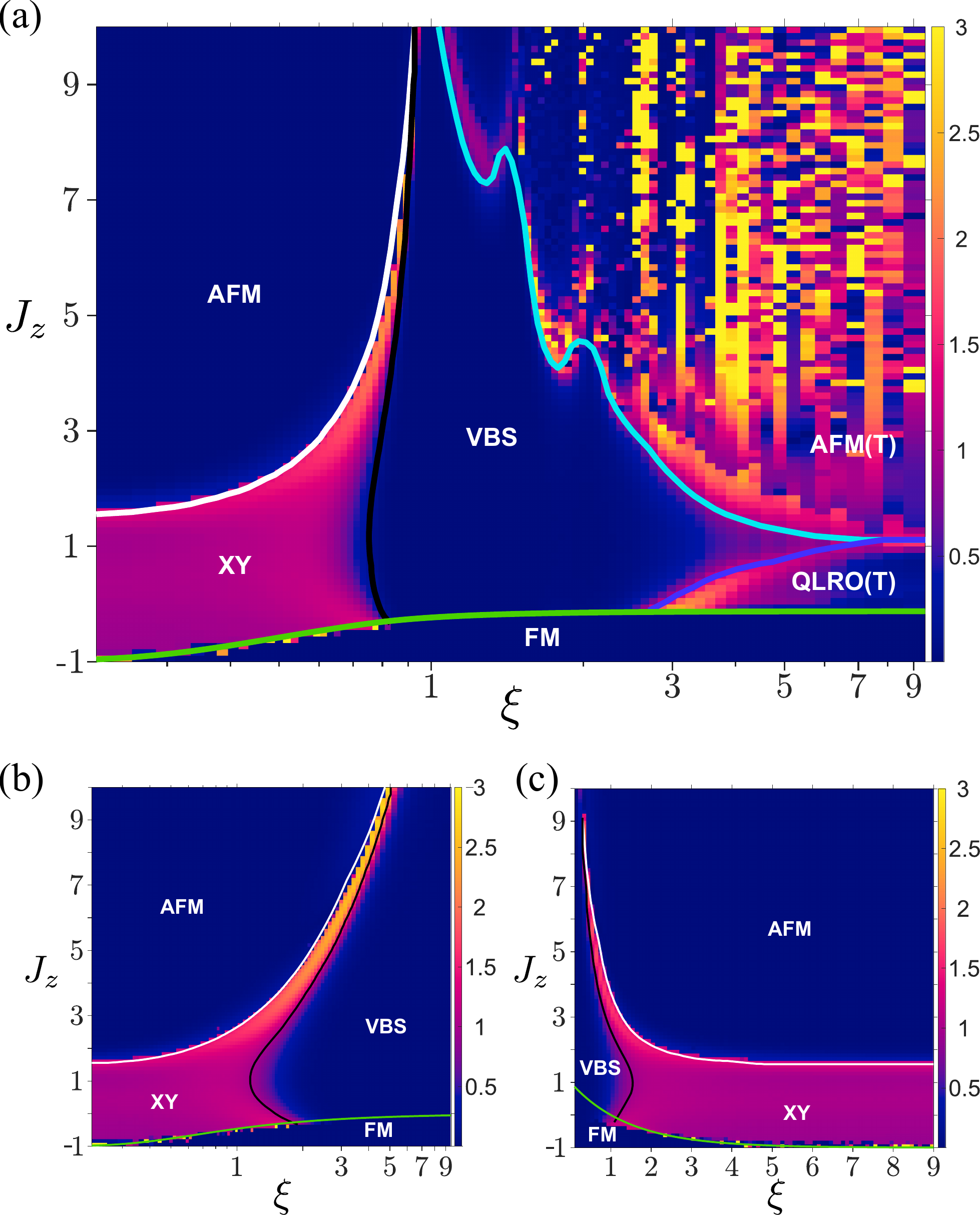}
		\caption{(a-c) Many-body phases of XXZ model $H_{XXZ}=\sum_{m>n}J(m-n)(S^+_mS^-_n/2+S^-_mS^+_n/2+J_zS^z_mS^z_n)$ for PLE interaction $J(x)=x\exp(-x/\xi)$ (a), exponential decay interaction $J(x)=\exp(-x/\xi)$ (b) and power-law decay interaction $J(x)=x^{-\xi}$ (c) based on the calculation of the effective central charge $c$. The phase boundary from FM phase to other phases is obtained via a spin-wave analysis Appendix\,\ref{suppD}. The XY-to-AFM and XY-to-VBS phase boundaries are numerically obtained by finding the place where $c$ starts to decrease $0.04$ below $c=1$. The VBS-to-AFM(T) and VBS-to-QLRO(T) phase boundaries are numerically obtained by finding the place where $c$ starts to increase from $c=1$.}\label{Phases_ceff}
	\end{figure}
\begin{figure}
		\centering
		\includegraphics[width=8.4cm]{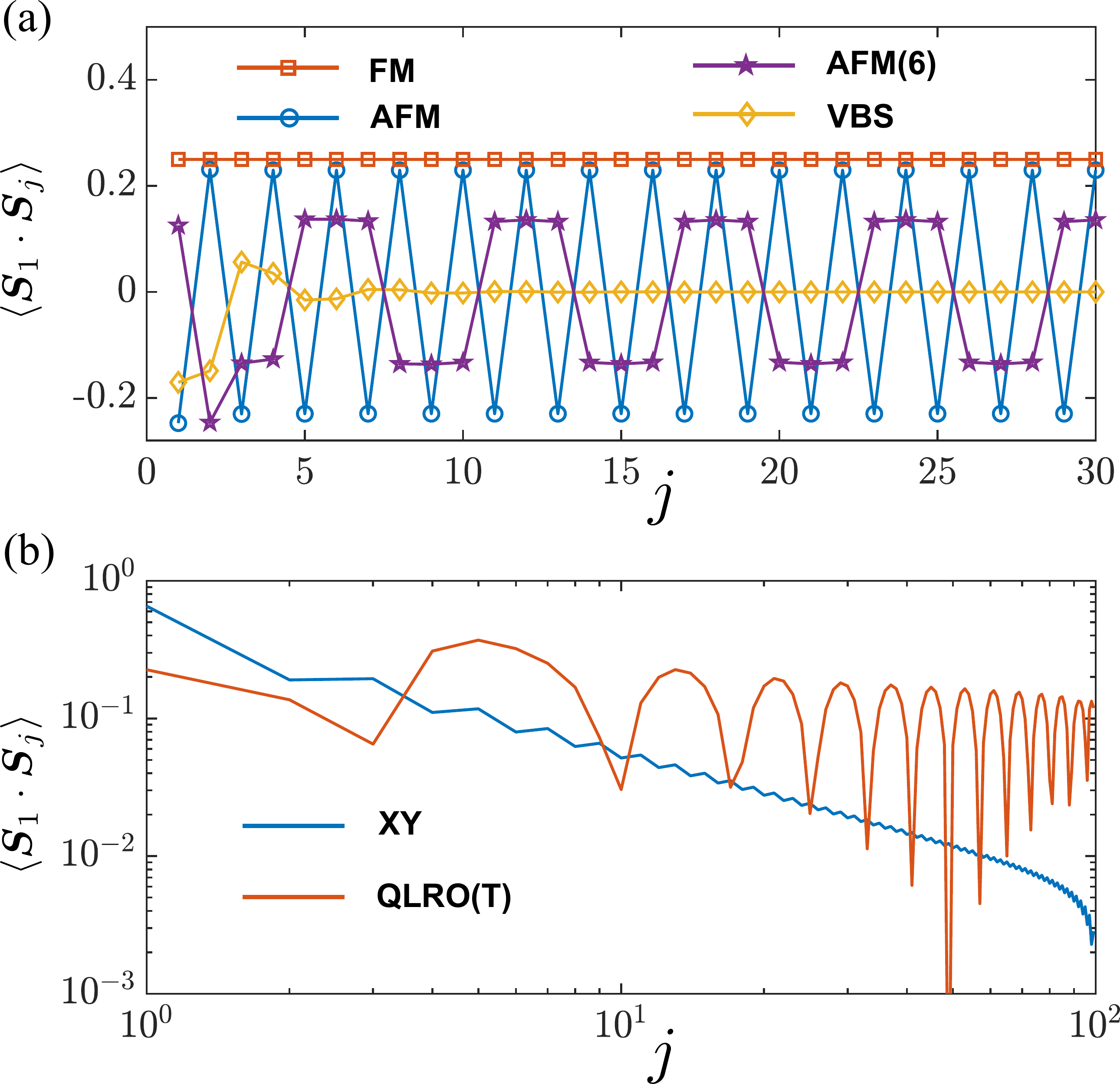}
		\caption{(a) Spin correlation deep in gapped phases. Here we present the AFM(T) phase with a specific period $T=6$. (b) Spin correlation deep in gapless phases.}\label{corrs}
	\end{figure}

	In order to explore the many-body phases, we numerically calculate the entanglement entropy and spin-spin correlation of the ground-state by using the DMRG method \cite{itensor}. The von Neumann entanglement entropy is defined as $S(\rho_A)=-\Tr(\rho_A\ln \rho_A)$ where $\rho_A$ is the reduced density matrix of the left $N_A$ cell of the chain. For a $1+1$-dimensional critical system, the entanglement entropy has a functional form (under open boundary condition)
	\begin{align}
		S(\rho_A)=\frac{c}{6}\log\qty(\frac{2N}{\pi}\sin(\frac{\pi N_A}{N}))+g+F.
	\end{align}
	Here $c$ is the central charge, $g$ is a constant and $F$ is a non-universal oscillating term\,\cite{Calabrese2004}. We extract the phase boundaries by numerically calculating the central charge of the ground state. After that, we identify each phase by the spatial profile of spin-spin correlation $\langle \boldsymbol{S}_i\cdot\boldsymbol{S}_j\rangle$ [see Fig.\,\ref{corrs}]. The phase diagram is displayed in Fig.\,\ref{Phases_ceff}(a). As a comparison, we also present the phase diagram of the XXZ models with exponential decay and power-law decay interactions [see Figs.\,\ref{Phases_ceff}(b-c)].
	
	From the phase diagrams shown in Fig.\,\ref{Phases_ceff}, it is found that four phases that are respectively antiferromagnetic (AFM), ferromagnetic (FM), XY and valence-bond-solid (VBS) phases occur for the PLE interaction when $\xi$ is not too large. We can understand qualitatively the physical origin of these phases by perturbatively truncating the long-range interaction to short-range interaction. Specifically, when $\xi\ll1$, only the nearest-neighbor interaction dominates due to the fast decay of the exponential term, i.e., $J(m-n)=J(1)\delta_{m,n+1}$. Therefore, the system is in the AFM (FM) phase for $\Delta>1$ $(\Delta<-1)$ and in the XY phase for $-1<\Delta<1$\,\cite{XXZ1,XXZ2}. As increasing $\xi$, the next-nearest-neighbor interaction becomes relevant, leading to $J(m-n)=J(1)\delta_{m,n+1}+J(2)\delta_{m,n+2}$. This system is know as the $J_1$-$J_2$ model\,\cite{j1j2,j1j2_2}, and it is geometrically frustrated due to the competition between the nearest-neighbor interaction $J(1)$ and the next-nearest-neighbor interaction $J(2)$. For $J_z=1$, this model supports a transition from $XY$ phase to the gapped VBS phase at the point $J(2)/J(1)\approx 0.2411$. For $J_z\gg0$ ($J_z\ll0$), one obtains an ordered antiferromagnetic (ferromagnetic) phase since the $zz$ interaction dominates. We stress that all the above phases also exist in the exponential and power-law decay interacting systems. The different forms of long-range interactions, however, modifies the boundaries between these phases.

	\begin{figure}
		\centering
		\includegraphics[width=8.4cm]{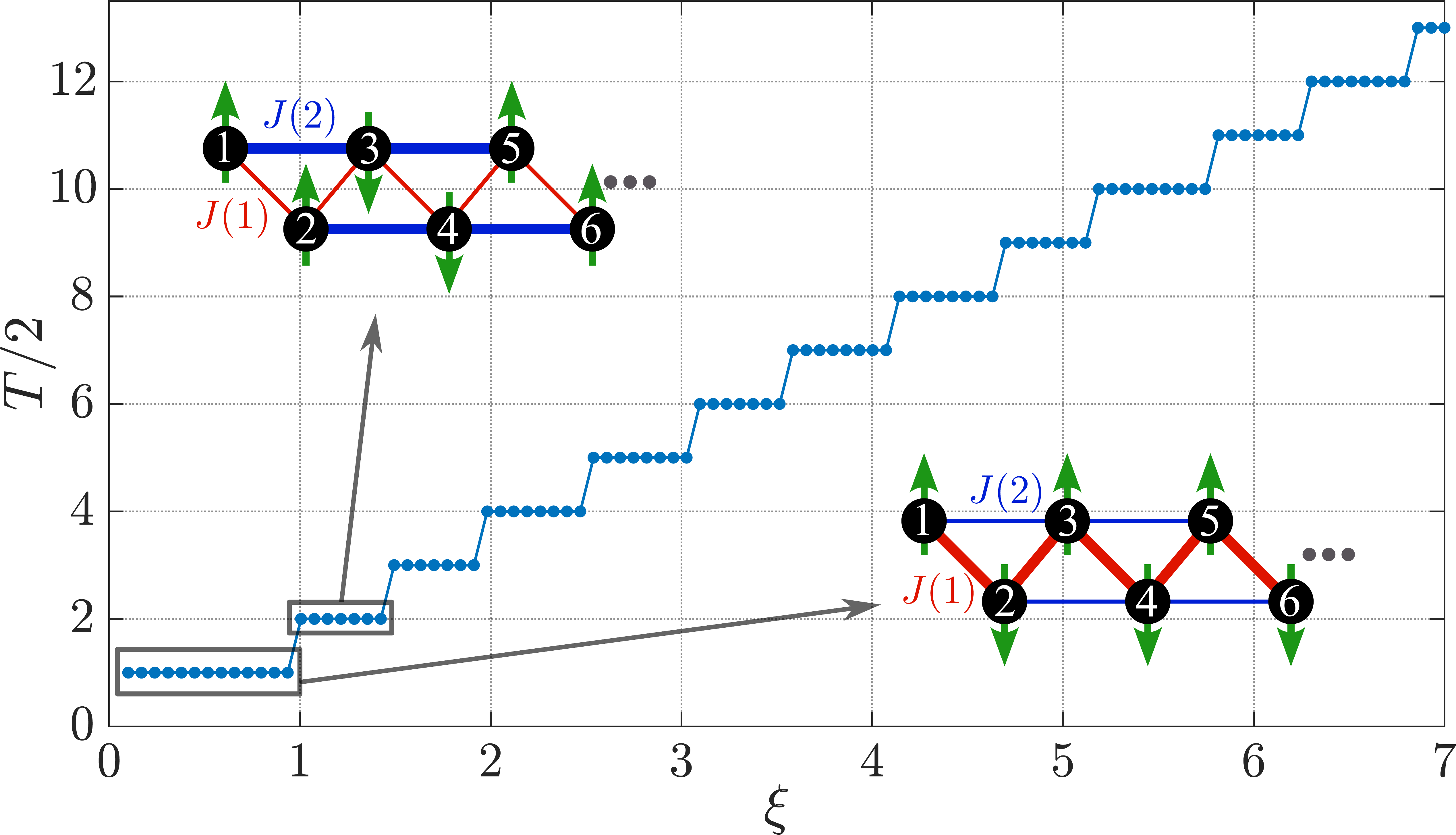}
		\caption{The period T versus the interaction length $\xi$. The result is obtained by a Monte Carlo simulation of the Ising model $H_{Ising}=\sum_{m>n}J(m-n)s_ms_n$. The chain size is $N=1500$. Insets are respectively the schematic configurations of the short-range Ising model in AFM(4) (top) and AFM (bottom) phase.}\label{AFM_T}
	\end{figure}
	
	We are more interested in the regime $\xi\gg1$ where the power-law factor of PLE interaction becomes relevant. Within this regime, two phases denoted respectively by AFM(T) and QLRO(T) appear only for PLE interaction. Let us firstly focus on the AFM(T) phase. A AFM(T) phase denotes an ordered, antiferromagnetic ground state with period T. For example, the AFM(6) phase is given by $|\uparrow \uparrow \uparrow\downarrow\downarrow\downarrow\uparrow\uparrow\uparrow\downarrow\downarrow\downarrow\cdots\rangle$. From Fig.\,\ref{Phases_ceff}, such AFM(T) phases exists in the strong $zz$ interaction limit, i.e., $J_z\ll1$. In this situation, the XXZ model is approximately given by $H_{XXZ}\approx \sum_{m>n}J(m-n)S^z_mS^z_n$. The quantum fluctuation of this model vanishes, therefore its ground state should coincide with the ground state of the classical Ising model $H_{Ising}=\sum_{m>n}J(m-n)s_ms_n$ where $s_m=\pm1$ represents the $m$-th classical spin. We then present the period T of the antiferromagnetic ground state in Fig.\,\ref{AFM_T}. It is found that the period increases with increasing the interaction length $\xi$. In order to understand the relation between T and $\xi$, one can consider again the short-range Ising model in which $J(m-n)=J(1)\delta_{m,n+1}+J(2)\delta_{m,n+2}$. When $\xi\ll1$, the nearest-neighbor interaction $J(1)$ dominates hence the spin pairs of the nearest neighbors tend to be anti-parallel, leading to the AFM phase. When the next-nearest-neighbor interaction $J(2)$ dominates, the spin pairs of the next-nearest neighbors tend to be anti-parallel, leading to the AFM(4) phase. In other word, the period T increases with increasing the geometry frustration, i.e., the ratio $J(2)/J(1)$. Keeping increasing $\xi$ integrates stronger and longer-range geometry frustration into the system hence the larger period T.
	
\begin{figure}
		\centering
		\includegraphics[width=8.4cm]{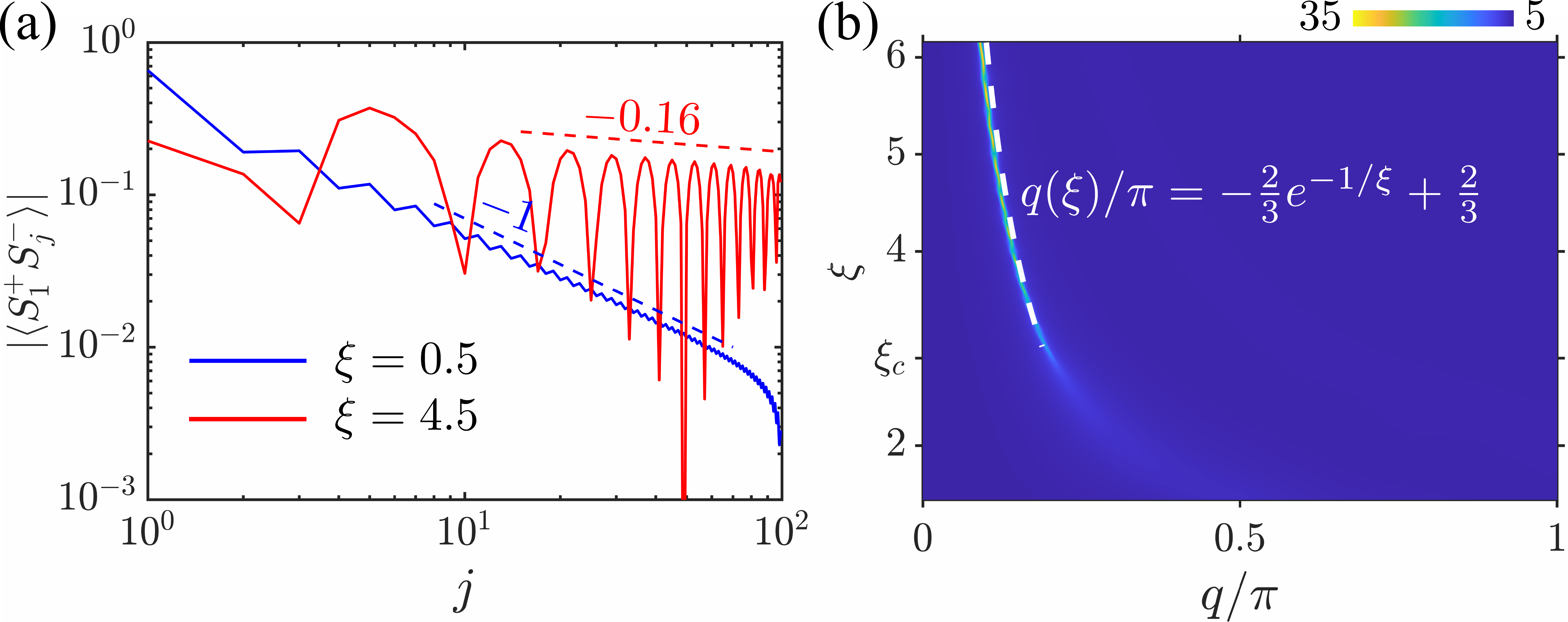}
		\caption{(a) Spin correlation for $\xi=0.5$ (blue solid line) and $\xi=4.5$ (red solid line). The system size is $N=100$. Dashed lines are fits $(j-1)^{\eta}$ with values of $\eta$ being labeled beside each curve. (b) Static structure factor $S_{xy}(q)$ versus the momentum $q$ and the interaction length $\xi$ for $N=300$. White dashed line is the fit $T(\xi)=3/(1-\exp(-1/\xi))$. The VBS-QLRO(T) phase transition point $\xi_c\approx 2.9$. In all plots, the $zz$ interaction strength is $J_z=0$.}\label{fig4}
	\end{figure}

	Turn to the QLRO(T) phase, for intermediate value of $J_z$, system undergoes a phase transition from the VBS phase to a critical phase. This critical phase exhibits quite distinctive behavior from the $XY$ phase appearing in short-range interacting limit. The spin correlation function $\langle S^+_iS^-_j\rangle\sim|i-j|^{-\eta}$ decays with a rather slow power law (e.g., $\eta\approx0.16$ at $\xi=4.5$ and $J_z=0$). Meanwhile, spin correlation is spiral with the incommensurate period T (T is not an integer) [see Fig.\,\ref{fig4}(a)]. Algebraic decay plus spiral behavior implies this critical phase is T-period quasi-long-range ordered phase (QLRO(T)). Besides, we find that the oscillation period T changes with interaction length. To characterize the relation between T and $\xi$, we further calculate the static structure factor in $x$-$y$ plane $S_{xy}(q)=\langle S^+(q)S^-(q)\rangle$, where $S^-(q)=\sum_{m=1}^N\exp(-iqx_m)S^-_m/\sqrt{N}$ denotes a spin density wave operator with momentum $q\in\{0,2\pi/N,4\pi/N,\cdots,(N-1)2\pi/N\}$. By numerically fitting the peaks of $S_{xy}(q)$, the approximate relation reads as $T(\xi)=3/(1-\exp(-1/\xi))$ for $J_z=0$ [see Fig.\,\ref{fig4}(b)].

\begin{figure}
		\centering
		\includegraphics[width=8.4cm]{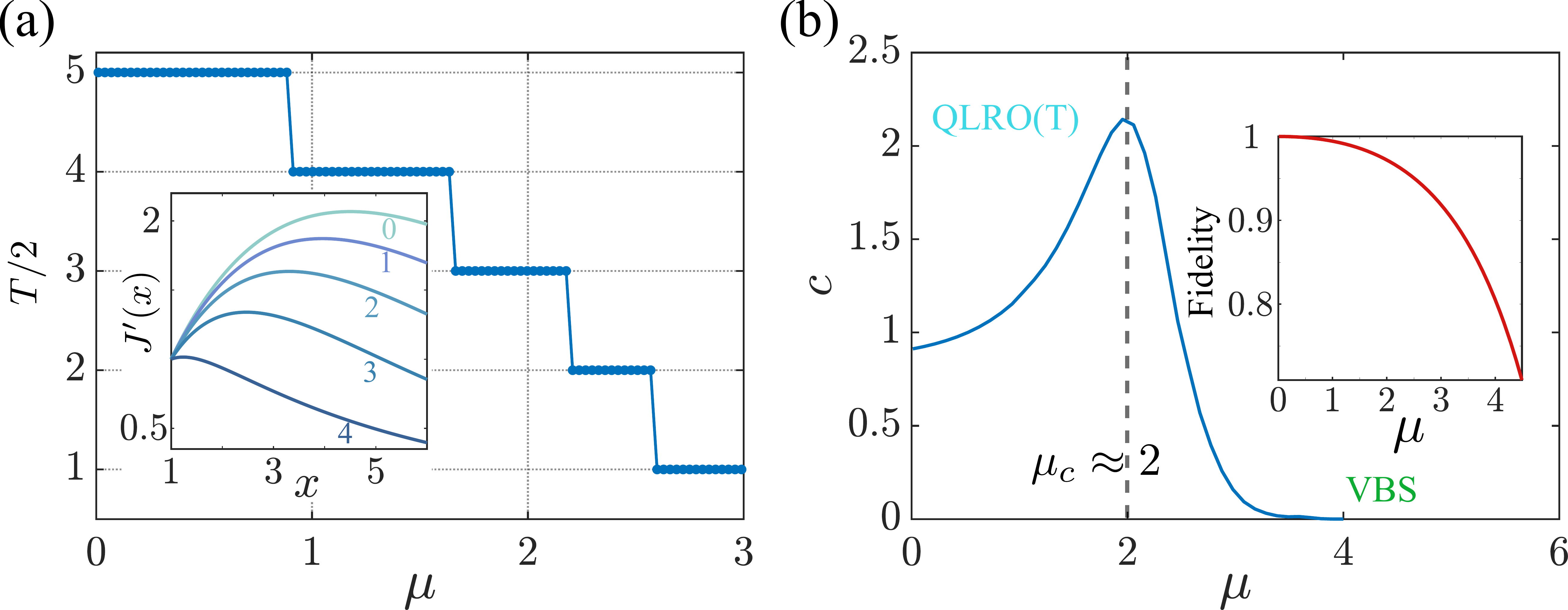}
		\caption{(a) Central charge for Hamiltonian $H_{XX}=\sum_{m>n}J'(m-n)(S^+_mS^-_n+S^+_nS^-_m)$ with $\xi_0=4.5$. Inset shows the fidelity of two normalized bound states for different values of $\mu$. (b) The period T of the ground state for the Ising model $H_{Ising}=\sum_{m>n}J'(m-n)s_ms_n$. The chain size is $N=1500$. Inset shows the spatial profile of $J'(x)$ with $\xi_0=3$ for different values of $\mu$. The values of $\mu$ are labeled besides each curves.}\label{PLE2EXP}
	\end{figure}

	Finally, we argue that the appearance of AFM(T) and QLRO(T) phases emerges from the strong geometry frustration brought by the power-law factor of PLE interaction, which explains their absence in exponential and power-law decay interacting systems. For the AFM(T) phase, our discussion based on the short-range Ising model explains the absence of AFM(T) in systems with exponential and power-law decay interactions: the next-nearest-neighbor interaction $J(2)$ is not able to dominate in comparison with $J(1)$, due to the intrinsicality of decay. Go beyond the short-range interaction approximation, we present the phase diagram of a system under the long-range interaction $J'(x)=(\exp(-x/\xi_0)-\exp(-x/\xi_1))/(\exp(-1/\xi_0)-\exp(-x/\xi_1))$ with $\xi_0$ being chosen within AFM(T) phase and $\xi_1=\xi_0-\mu$. The parameter $\mu$ can be considered as the inverse strength of geometry frustration. For $\mu\to0$, $J'(x)$ reduces to PLE interaction with interaction length $\xi_0$. As the increase of $\mu$, the power-law growth of PLE interaction weakens [see the inset of Fig.\,\ref{PLE2EXP}(a)]. For $\mu\to\xi_0$, $J'(x)$ reduces to exponential interaction with interaction length $\xi_0$. The result in Fig.\,\ref{PLE2EXP}(a) demonstrates that the period T decreases with increasing $\mu$, which means T decreases with decreasing geometry frustration. This is also consistent with the statement in Fig.\,\ref{AFM_T}. For the QLRO(T) phase, we present the phase diagram for a system $H_{XX}=\sum_{m>n}J'(m-n)(S^+_mS^-_n+S^+_nS^-_m)$ with $\xi_0$ being chosen within QLRO(T) phase in Fig.\,\ref{PLE2EXP}(b). As expected, system undergoes a QLRO(T)-VBS phase transition when $\mu>\mu_c\approx2$, which supports our above argument. At the transition point, the distinguishability between $J'(x)$ and PLE interaction with interaction length $\xi_0$ is approximate $0.97$. Here we define the distinguishability of two kind of interactions $J_{1,2}(x)$ as the fidelity of two normalized dressed bound states $|\langle\varphi_{1}|\varphi_{2}\rangle|$ based on $J_1(x)\propto\langle x|\varphi_{1}\rangle$ and $J_2(x)\propto\langle x|\varphi_{2}\rangle$. This implies that the spiral phase only occurs when the interaction are highly overlapped with PLE interaction.

	\section{Discussion of experimental implementation of PLE interaction}\label{IV}

	Regarding the experimental implementation of our proposal, the candidates include the system of color centers or cold atoms integrated into 1D photonics crystal waveguide\,\cite{douglas2015quantum,Mdlukin1,Mdlukin2,topo2d}, and the circuit QED system where the superconducting transmon qubits are coupled to an superconducting metamaterial waveguide\,\cite{KimTopological,topophoton,RMP87,RMP90}. The photonic analog of SSH model has been experimentally realized both in photonic crystal waveguide\,\cite{keil,xiaomeng,chen} and superconducting metamaterials\,\cite{lee2018topolectrical,topocircut}. As an example, let us choose the coupled circuit QED array to discuss the realization of the unconventional PLE interaction based on our proposal. Since the most important element to implement PLE interaction is the longer-range hopping terms beyond nearest-neighbor of bath, we mainly focus on the potential implement of long-range interactions of bath.
	
\begin{figure}
		\centering
		\includegraphics[width=8.4cm]{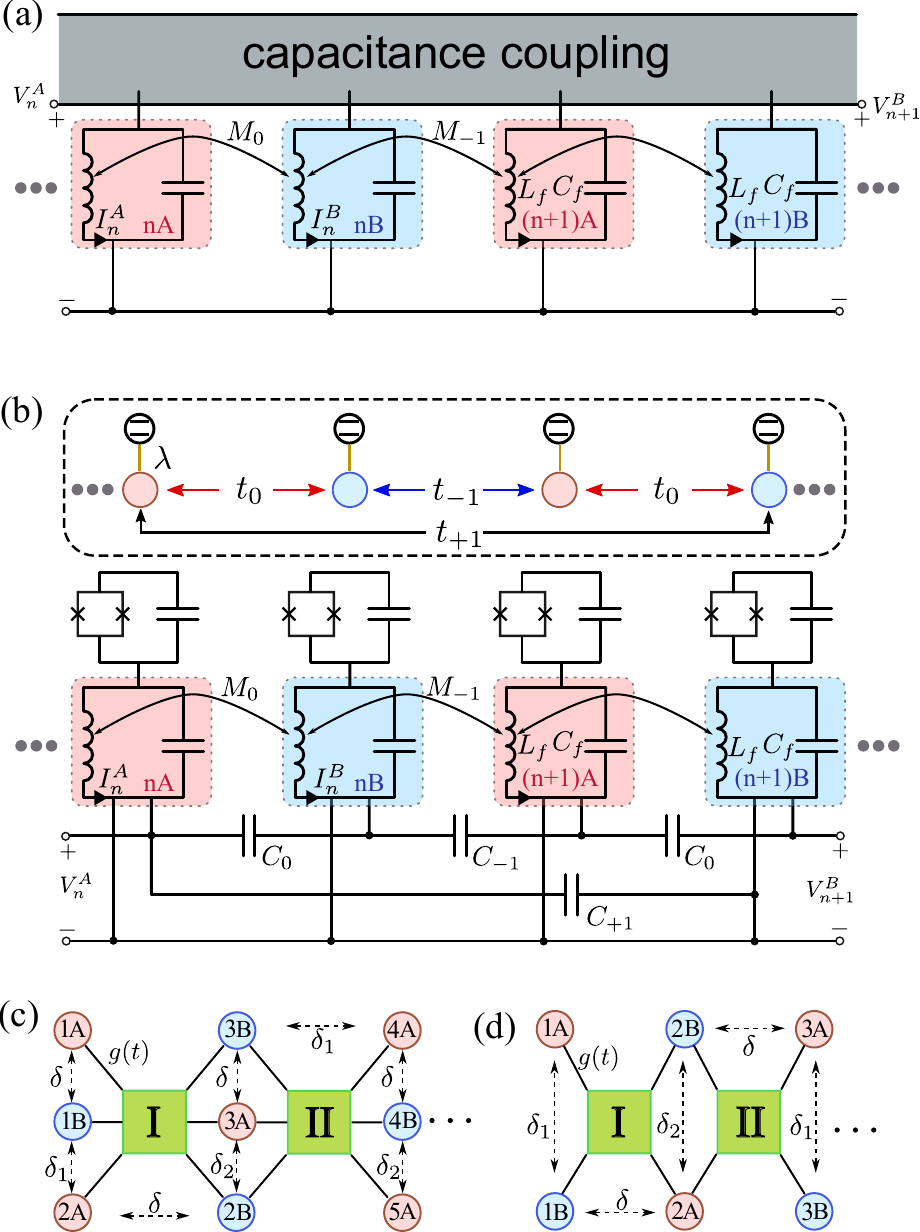}
		\caption{(a) Full circuit model used in the simulations of bath endowed with arbitrary configuration. The dashed area denotes the coupling between different sites depending on the requirement of bath configuration. For example, if sites A of unit cell $j+m$ and B of unit cell $j$ are required by coupling, then they will be connected by a capacitance denoted by $C_{j+m}$. (b) Full circuit model used in the simulations of Eq.\,(\ref{eq1}) with bath configuration shown in Fig.\,\ref{fig1}(b) of main text, we assume that each site couples an emitter, the parameters are further discussed in the text. (c-d) Possible circuit QED architectures to implement the bath Hamiltonian in Fig.\,\ref{fig1}(b) (see (b)) and Fig.\,\ref{fig1}(c) (see (c)) of main text.  A set of LC resonators with frequency $\omega_{j,\alpha}$ is coupled in groups to auxiliary resonators (denoted by Roman numbers I, II ...) with frequency $\omega_{\rm aux}$ and time-dependent coupling $g(t)$. The bare energies of lattice resonator have a distribution of energy difference denoted by $\delta,\delta_1,\delta_2$.}\label{fig7}
	\end{figure}

	We firstly derive a generic bath Hamiltonian $H_b$ based on the experimental proposal in\,\cite{KimTopological}. The resulting hopping configuration of $H_b$ can be arbitrary and $H_b$ restores the chiral symmetry $\sigma_z$. To this end, we consider the array of coupled LC resonators [see Fig.\,\ref{fig7}(a)]. Sites A of unit cell $j+m$ and B of unit cell $j$ are coupled by the capacitance $C_{j+m}$ and the mutual inductance $M_{j+m}$. The flux variable of each node is $\Phi^{A/B}_j=\int_{-\infty}^t V^{A/B}_j(t')dt'$ and the current going through each inductor is $I^{A/B}_j$. Then the Lagrangian is
	\begin{equation}
		\begin{aligned}
			\mathcal{L}=&\sum_j^{N}\frac{C_f}{2}\qty[(\dot{\Phi}^A_j)^2+(\dot{\Phi}^B_j)^2]-\frac{L_f}{2}\qty[\qty(I^A_j)^2+\qty(I^B_j)^2] \\
			& + \frac{C_0}{2}(\dot{\Phi}^A_j-\dot{\Phi}^B_j)^2-M_0I^A_jI^B_j-M_{-1}I^A_{j+1}I^B_j \\
			& + \sum_{m=1}^PC_{-m}(\dot{\Phi}^A_{j+m}-\dot{\Phi}^B_j)^2 + \sum_{n=1}^QC_{+n}(\dot{\Phi}^A_{j-n}-\dot{\Phi}^B_j)^2
		\end{aligned}.
	\end{equation}
	The node flux variables are given by
	\begin{equation}
		\begin{aligned}
			&\Phi^A_j=L_fI^A_j+M_0I^B_j+M_{-1}I^B_{j-1},\\ &\Phi^B_j=L_fI^B_j+M_0I^A_j+M_{-1}I^A_{j+1}.
		\end{aligned}
	\end{equation}
	The Lagrangian can be rewritten with the Fourier transformation $\Phi^{A/B}_j=\sum_k\exp(ijk)\Phi^{A/B}_k/\sqrt{N}$ and $I^{A/B}_j=\sum_j\exp(ijk)I^{A/B}_k/\sqrt{N}$, which leads to
	\begin{equation}
		\begin{aligned}
			\mathcal{L}=&\sum_k \frac{C_t}{2}(\dot{\Phi}^A_{-k}\dot{\Phi}^A_k+\dot{\Phi}^B_{-k}\dot{\Phi}^B_k)-C_{\lambda}(k)\dot{\Phi}^A_{-k}\dot{\Phi}^B_k \\
			&-\frac{L_f(\Phi^A_{-k}\Phi^A_k+\Phi^B_{-k}\Phi^B_k)-M_{\lambda}(k)\Phi^A_{-k}\Phi^B_{k}}{2L^2_{\lambda}(k)}
		\end{aligned},
	\end{equation}
	where $C_t=C_f+C_0+\sum_{m=1}^PC_{-m} + \sum_{n=1}^QC_{+n}$, $C_{\lambda}(k)=C_0+\sum_{m=1}^PC_{-m}e^{-imk} + \sum_{n=1}^QC_{+n}e^{ink}$ and $M_{\lambda}(k)=M_0+M_{-1}e^{-ik}$. Then the Hamiltonian is given by $H_b=\sum_{k,\alpha\in\qty{A,B}}Q^{\alpha}_k\dot{\Phi}^{\alpha}_k-\mathcal{L}=\sum_{k}\mqty(Q^A_{-k} & Q^{B}_{-k})\mathcal{M}_{Q}(k)\mqty(Q^A_{k} & Q^{B}_{k})^T/2+\mqty(\Phi^A_{-k} & \Phi^{B}_{-k})\mathcal{M}_{\Phi}(k)\mqty(\Phi^A_{k} & \Phi^{B}_{k})^T/2$ with
	\begin{align}
		M_{Q}(k)=\frac{1}{C^2_{\rm tot}(k)}\mqty(C_t & C_{\lambda}(-k) \\ C_{\lambda}(k) & C_t)
	\end{align}
	and
	\begin{align}
		M_{\Phi}(k)=\frac{1}{L^2_{\rm tot}(k)}\mqty(L_f & -M_{\lambda}(k) \\ -M_{\lambda}(-k) & L_f).
	\end{align}
	Here $Q^{A/B}_k=\partial\mathcal{L}/\partial\dot{\Phi}^{A/B}_k$, $C^2_{\rm tot}(k)=C^2_t-C_{\lambda}(k)C_{\lambda}(-k)$ and $L^2_{\rm tot}(k)=L^2_0-M_{\lambda}(k)M_{\lambda}(-k)$. We then can impose the transformation
	\begin{equation}
		\begin{aligned}
			\hat{a}_{k,\alpha}=\frac{1}{\sqrt{2\hbar}}\qty(\frac{\hat{\Phi}^{\alpha}_k}{\sqrt{Z(k)}}+i\sqrt{Z(k)}\hat{Q}^{\alpha}_k)
		\end{aligned}
	\end{equation}
	after the canonical quantization $[\hat{\Phi}^{\alpha}_k,\hat{Q}^{\alpha}_{k'}]=i\hbar\delta_{\alpha,\beta}\delta_{k,k'}$. Here $\hat{a}_{k,\alpha}$ is the annihilation operator that obeys the commutation relations $[\hat{a}_{k,\alpha},\hat{a}^{\dagger}_{k',\beta}]=\delta_{\alpha,\beta}\delta_{k,k'}$ and $[\hat{a}_{k,\alpha},\hat{a}_{k',\beta}]=[\hat{a}^{\dagger}_{k,\alpha},\hat{a}^{\dagger}_{k',\beta}]=0$, and $Z(k)=\sqrt{C_{t}L^2_{\rm tot}(k)/L_fC^2_{\rm tot}(k)}$. Under these notations, $H_b$ is given by $H_b=H_{b0}+V$ where $H_{b0}=\sum_k \omega_0(k)(\hat{a}^{\dagger}_{k,A}\hat{a}_{k,A}+\hat{a}_{-k,A}\hat{a}^{\dagger}_{-k,A}+\hat{a}^{\dagger}_{k,B}\hat{a}_{k,B}+\hat{a}_{-k,B}\hat{a}^{\dagger}_{-k,B})/2$ and
	\begin{equation}
		\begin{aligned}
			V=&\sum_k \mqty(a^{\dagger}_{k,A} & a^{\dagger}_{k,B})\mqty(0 & \tilde{h}(k) \\ \tilde{h}^*(k) & 0)\mqty(a_{k,A} \\ a_{k,B}) \\ &+\mathrm{counterrotating\ terms}
		\end{aligned},
	\end{equation}
	with $\omega_0(k)=(C_{\rm tot}^2(k)L^2_{\rm tot}(k)/C_tL_f)^{-1/2}$ and $\tilde{h}(k)=C_{\lambda}(k)\omega_0(k)/2C_t-M_{\lambda}(k)\omega_0(k)/2L_0$ (we set $\hbar=1$). Under the weak coupling approximations $C_0+\sum_{m=1}^PC_{-m}+\sum_{n=1}^QC_{+n}\ll C_f$ and $|M_0|+|M_{-1}|\ll L_f$, both the $k$-dependence of $\omega_0(k)$ and the counterrotating terms of $V$ can be neglected, which results in $H_b=\sum_{k}\mqty(\hat{a}^{\dagger}_{k,A} & \hat{a}^{\dagger}_{k,B})H_b(k)\mqty(\hat{a}_{k,A} & \hat{a}_{k,B})$ with
	\begin{align}
		H_b(k)=\mqty(\omega_0 & h(k) \\ h^*(k) & \omega_0).
	\end{align}
	Here $\omega_0=1/\sqrt{L_fC_{t}}$ and
	\begin{align}
		h(k)=\frac{\omega_0}{2}\qty(\frac{C_{\lambda}(k)}{C_{t}}-\frac{M_{\lambda}(k)}{L_f}).
	\end{align}
	Together with the bath configuration used in Eq.\,(\ref{eq1}), we have
	\begin{equation}
		\begin{aligned}
			&t_{m}=\frac{\omega_0}{2}\qty(\frac{C_m}{C_t}+\frac{M_m}{L_f}),\ m=0,-1 \\
			&t_{n}=\frac{\omega_0}{2}\frac{C_m}{C_t},\ n\ne0,-1
		\end{aligned}.
	\end{equation}
	
\begin{figure}
		\centering
		\includegraphics[width=7.9cm]{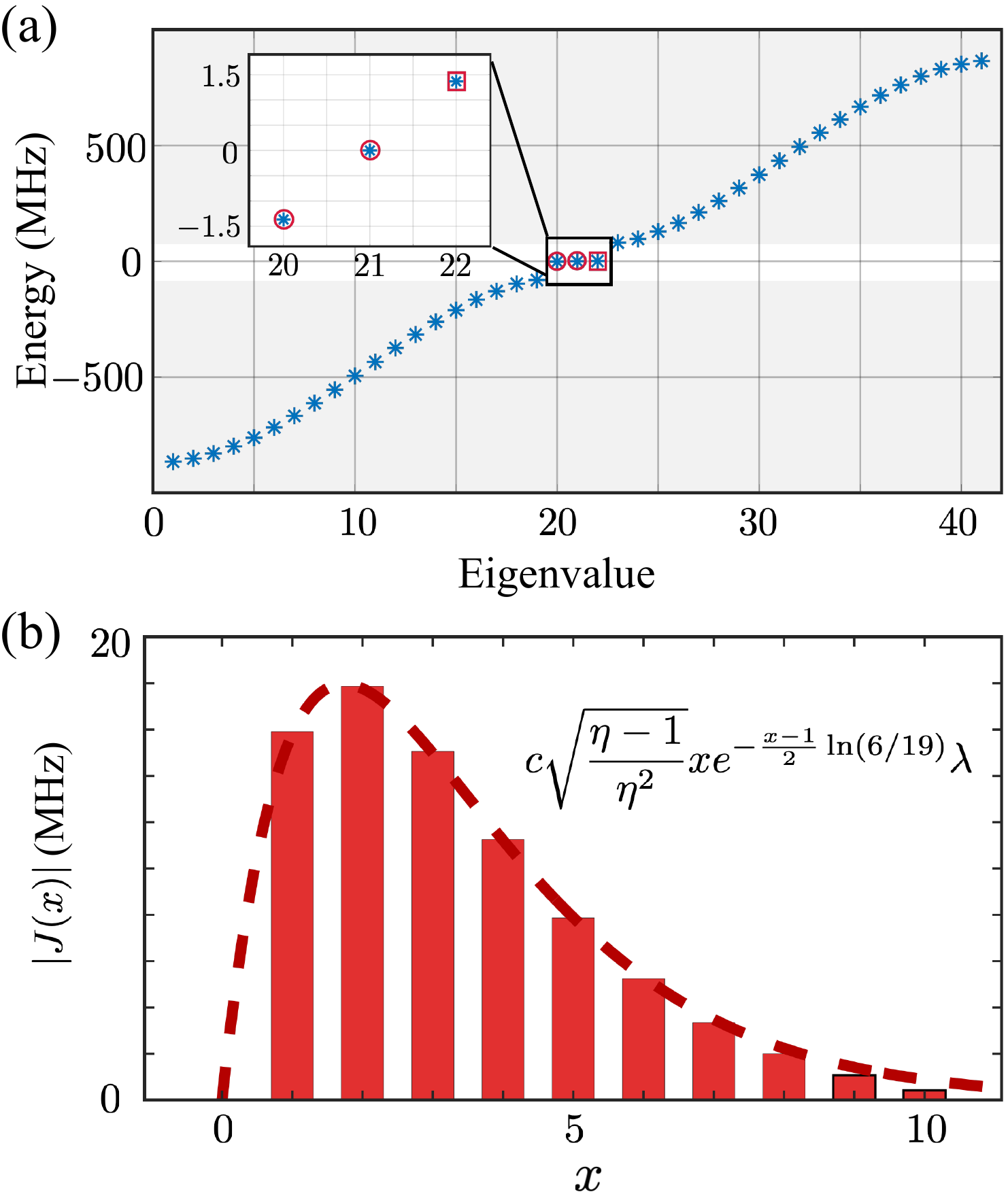}
		\caption{(a) Schematic of QEs-waveguide model. Eigenspectrum of system in single-excitation regime, where we set $\omega_q=\omega_0=0$ for simplicity. The bath Hamiltonian under periodic condition is $H_b(k)=d_x(k)\sigma_x+d_y(k)\sigma_y$ with $h(k)/t_0=(e^{ik}+\sqrt{6/19})^2/e^{ik}$. The encircled stars represent edge state energies, and the dressed bound state energy is marked by the rectangle. (b) Spatial profile of d-d interaction. The histogram and dashed line correspond to the numerical solution and analytical prediction. Here $\eta\approx 1.26$, and the correction factor due to the open boundary condition is numerically fixed by $c\approx 0.8$.}\label{fig8}
	\end{figure}

	Based on the above results, the minimal model exhibiting PLE interaction, where $h(k)=t_0+t_{-1}e^{-ik}+t_{+1}e^{ik}$ can be realized. To show this, consider the parameters $C_{f},C_{0},C_{-1},C_{+1},M_{0},M_{-1},L_{f}=$ 253 fF, 46 fF, 41 fF, 13 fF, -1 pH, -1 pH, 1.9 nH. Given the values of these circuit elements, the corresponding parameters of the topological waveguide QED system displayed in Fig.\,\ref{fig7}(b) can be obtained as $\omega_0\approx 6.14$ GHz, $t_0\approx 399$ MHz, $t_{+1}\approx 113$ MHz and $t_{-1}\approx 355$ MHz. Since the d-d interaction can be recovered by the single QE dressed bound state according to $J(x)\propto \lambda \langle x|\varphi_E\rangle$, we assume that only one QE is resonantly coupled to bath in sublattice A at the middle of chain, with chain size $L=20$ and coupling strength $\lambda/2\pi=90$ MHz. The spectrum of system contains 41 eigenvalues in single-excitation regime. Except the bulk state energies, two in-gap energies $\approx -1.37,0$ MHz corresponding to the edge states of bath appear [see the red circles in Fig.\,\ref{fig8}(a)]. Moreover, due to the finite size bath, the dressed bound state energies $\approx 1.37$ MHz [see the red square in Fig.\,\ref{fig8}(a)] is slightly deviated from the analytical prediction $E=0$. In spite of the small size bath, the obtained PLE interaction [see Fig.\,\ref{fig8}(b)] by numerically calculation consist with the analytical prediction up to a constant correction. This correction mainly originates from the non-zero occupation of dressed bound state in sublattice A, due to the open boundary condition in small bath size. Therefore, our results show that the PLE interaction can be implemented in the current experimental platform in small bath size.
	
	Besides the above circuit proposal, another way to implement the long-rang hopping terms of bath is reported in\,\cite{Vega}. The basic idea is that the LC resonators representing lattice sites are coupled in groups to auxiliary resonators according to
	\begin{align}
		H_{\rm tot}=\sum_{j,\alpha}^N\omega_{j,\alpha}a^{\dagger}_{j,\alpha}a_{j,\alpha}+\sum_{\mu}\omega_{\rm aux}b^{\dagger}_{\mu}b_{\mu}+V(t)
	\end{align}
	with
	\begin{align}
		V(t)=\sum_{\mu=\rm 1,2,\cdots}\sum_{\langle j,\alpha\rangle_\mu}g(t)a^{\dagger}_{j,\alpha}b_{\mu}+h.c..
	\end{align}
	Here $a_{j,\alpha}$ is the annihilation operator of the bath mode corresponding to the $\alpha$ sublattice in $j$-th cell, $b_{\mu}$ denotes the annihilation of $\mu$-th auxiliary cavity mode, endowed with the bare frequency $\omega_{\rm aux}$. The bare frequency of lattice mode $\omega_{j,\alpha}$ is correlated to its nearest neighbor according to the relation $\omega_{j,B}=\omega_{j,A}+d_{j,\rm intra}$ and $\omega_{j,A}=\omega_{j-1,B}+d_{j-1,\rm inter}$, where $d_{j,\rm intra}$ and $d_{j,\rm inter}$ being the detunings. In the following derivation, we assumes the detunings $\qty{d_1,d_2,\cdots}$ are given by [see Figs.\,\ref{fig7}(c-d)] either
	\begin{equation}
		\begin{aligned}
			d_{j,\rm intra}=\delta,\ 
			d_{j,\rm inter}=\left\{
			\begin{aligned}
				&\delta_1,\!\!\mod(j,2)=1, \\
				&\delta_2,\!\!\mod(j,2)=0, \\
			\end{aligned}
			\right.
		\end{aligned}
	\end{equation}
	or
	\begin{equation}
		\begin{aligned}
			d_{j,\rm inter}=\delta,\ 
			d_{j,\rm intra}=\left\{
			\begin{aligned}
				&\delta_1,\!\!\mod(j,2)=1, \\
				&\delta_2,\!\!\mod(j,2)=0, \\
			\end{aligned}
			\right.
		\end{aligned}
	\end{equation}
	depending on the bath configuration. Further, the time-dependent coupling is $g(t)\propto\sum_{m}A_m\cos(\Omega_mt)$. The notation $\langle j,\alpha\rangle$ denotes the summation over the sites connected by $\mu$-th auxiliary cavity. In the rotating frame with respect to $\sum_{j,\alpha}^N\omega_{j,\alpha}a^{\dagger}_{j,\alpha}a_{j,\alpha}+\sum_{\mu}\omega_{\rm aux}b^{\dagger}_{\mu}b_{\mu}$, we have
	\begin{align}
		\tilde{V}(t)=\sum_{\mu=\rm I,II,\cdots}\sum_{\langle j,\alpha\rangle_\mu}g(t)a^{\dagger}_{j,\alpha}b_{\mu}e^{i(\omega_{j,\alpha}-\omega_{\rm aux})}+h.c..
	\end{align}
	Then, under weak coupling approximation $|\omega_{j,\alpha}-\omega_{\rm aux}|\gg|\omega_{j,\alpha}-\omega_{i,\beta}|\gg |g(t)|$, the long-range hopping terms of lattice bath can be obtained by adiabatically eliminating the auxiliary cavity as
	\begin{equation}
		\begin{aligned}
			&H_{\rm eff}\approx \frac{1}{2}\sum_{m,n}\frac{A_mA_n}{4}\qty(e^{i(\Omega_m+\Omega_n)t}+e^{i(\Omega_m-\Omega_n)t}+c.c.)\times \\ &\sum_{\mu,\langle j,\alpha\rangle_{\mu}\langle i,\beta\rangle_{\mu}}\!\! \omega^{i,j}_{\alpha,\beta}a^{\dagger}_{j,\alpha}a_{i,\beta}e^{-i(\omega_{j,\alpha}-\omega_{i,\beta})t}+h.c.,
		\end{aligned}
	\end{equation}
	where $\omega^{i,j}_{\alpha,\beta}=(\omega_{\rm aux}-\omega_{j,\alpha})^{-1}\!+\!(\omega_{\rm aux}-\omega_{i,\beta})^{-1}$. Then, one can obtain the desired long-range hopping by taking the frequencies satisfying the resonant conditions $\Omega_m\pm\Omega_n=(\pm)(\omega_{j,\alpha}-\omega_{i,\beta})$, while the undesired hopping terms with the off-resonant frequencies will be greatly suppressed due to the weak coupling approximation.
	
	For the bath configuration in Fig.\,\ref{fig1}(b) of main text, we couple six lattice sites in groups to the auxiliary resonators, and with the energy distribution shown in Fig.\,\ref{fig7}(b). One can chose the frequencies according to (consider the first auxiliary cavity and coupled sites for simply)
	\begin{equation}
		\begin{aligned}
			&\Omega^+_{1,2}=\Delta^{1,A}_{2,B},\ \Omega^+_{2,3}=\Delta^{2,A}_{3,B},\ \Omega^+_{4,5}=\Delta^{1,A}_{1,B}, \\ & \Omega^+_{4,6}=\Delta^{1,B}_{2,A},\ \Omega^+_{5,6}=\Delta^{2,B}_{3,A},\ \Omega^+_{1,3}\ne\Delta^{1,B}_{3,A},
		\end{aligned}
	\end{equation}
	where $\Omega^{+}_{i,j}=\Omega_i+\Omega_j$ and $\Delta^{j,\alpha}_{i,\beta}=\omega_{i,\beta}-\omega_{j,\alpha}$. Under these conditions, only the hopping terms $a^{\dagger}_{m,A}a_{m,B}+h.c.$ for $m=1,2,3$ with the hopping strength $t_0$, the hopping terms $a^{\dagger}_{m,B}a_{m+1,A}+h.c.$ for $m=1,2,3$ with the hopping strength $t_{-1}$ and the hopping terms $a^{\dagger}_{m,A}a_{m+1,B}+h.c.$ $m=1,2$ with the hopping strength $t_{+1}$ are resonant, while others are off-resonant. Finally, the desired hopping terms with certain strengths $\qty{t_0;t_{-1};t_{+1}}$ can be implemented by taking $A_1A_3=t_{+1},\ A_4=A_5=\sqrt{t_{0}},\ A_6=t_{-1}/t_0$.
	
	For PLE interaction without chirality, i.e., the bath configuration in Fig.\,\ref{fig1}(c) of main text, the set-up is different from the former case. Here we couple four lattice sites in groups to the auxiliary resonators, and with the energy distribution shown in Fig.\,\ref{fig7}(c). A possible resonant condition reads as
	\begin{align}
		\Omega^+_{1,2}=\Delta^{1,A}_{2,A},\ \Omega^+_{2,3}=\Delta^{1,B}_{2,B},\ \Omega_4=\Delta^{1,B}_{2,A}, \Omega^+_{1,3}\ne\Delta^{1,A}_{1,B}
	\end{align}
	associating with the amplitudes condition $A_1A_2=t'_0=A_2A_3$ and $A_4=t'_1$ to obtain the desired hopping structure shown in main text.

	\section{Conclusion}\label{V}
	In summary, we have shown how the PLE interaction emerges in a 1D waveguide QED system by appropriately designing the structure of lattice endowed with chiral symmetry. Applying the PLE interaction to a spin many-body model, we uncovered the exotic many-body phases in this system for the first time. Especially, we demonstrated the emergence of two spiral phases as the consequence of power-law factor for PLE interaction, which has no counterpart for other type of long-range interaction.
	
	Besides those studied here, firstly, we expect our work could stimulate further studies for the unconventional long-range interactions by engineering the topological properties of photonic lattice. Secondly, our work demonstrates the topological waveguide QED system provides a platform of quantum simulation to discover novel complex phases of matter, which may further be explored beyond the XXZ model. Thirdly, it would be an interesting outlook to investigate how this kind of interaction affects the propagation of quantum information, namely the Lieb-Robinson bounds\,\cite{Hauke,Richerme,Jurcevic}. And lastly, it is worth studying the topological properties of PLE interacting system since long-range interaction could alter the topological intrinsicality of matters\,\cite{Davide,Kristian,gzp,Nick}.
	
	We thank Prof. Tao Shi for valuable discussions. This work is supported by the National Key Research and Development Program of China grant 2021YFA1400700. The computing work in this paper is supported by the Public Service Platform of High Performance Computing provided by Network and Computing Center of HUST.

	\appendix
	\section{Exact solution of the Edge state}\label{suppA}
	To support the analytical PLE d-d interactions given in Eq.\,(\ref{int}) of the main text, we provide the detailed derivation for the chiral edge state of bath in subsection A. We also show how to implement nonchiral edge state in subsection B to support the interaction used in the simulation of many-body phase.
	
	\subsection{Edge state with chirality}
	Our goal is to solve the eigenequation in real space
	\begin{align}
		\qty[H_b+\epsilon\ketbra{x_{1,A}}{x_{1,A}}]\ket{\varphi_E}=0\ket{\varphi_E},
	\end{align}
	where $\epsilon=\lambda^2/(E_q-\omega_q)\rightarrow\infty$ for $E_q=\omega_q=\omega_0=0$ is the effective potential induced by the QE. Inserting $H_b=d_x(k)\sigma_x+d_y(k)\sigma_y$ into this equation, we have
	\begin{subequations}
		\begin{equation}
			\begin{aligned}\label{eq32}
				\sum_q\Big[\sum_{m}^{P}&t_{-m}\braket{\qty(q-m)_B}{\varphi_E}+\sum_{n}^{Q}t_{+n}\braket{\qty(q+n)_B}{\varphi_E}\\ &+\epsilon\delta_{q,x_1}\braket{q_A}{\varphi_E}\Big]\!=\!0,
			\end{aligned}
		\end{equation}
		\begin{align}
			&\sum_q\left[\sum_{m}^{P}t^*_{-m}\braket{\qty(q+m)_A}{\varphi_E}\!+\!\sum_{n}^{Q}t^*_{+n}\braket{\qty(q-n)_A}{\varphi_E}\right]\!=\!0,
		\end{align}
	\end{subequations}
	where $q\in \mathbb{Z}$ is the cell index. To solve this difference equation, we perform a Z-Transform, and obtain
	\begin{subequations}\label{eq33}
		\begin{align}
			&h(z)\Phi_B(z)+\epsilon\braket{x_{1,A}}{\varphi_E}z^{-x_{1,A}}=0, \\
			&h^*(z^{-1})\Phi_A(z)=0,
		\end{align}
	\end{subequations}
	where $h(z)\coloneqq h(k\to -i\ln z)$. One immediate result is $\Phi_A(z)=0$, which implies that the edge state component on sublattice A vanishes. The expression of $h(z)$ can be rewritten as a compact form $h(z)=f(z)/g(z)$ with $f(z)=\sum_{m}^{P}t_{-m}z^{P-m}+\sum_{n}^{Q}t_{+n}z^{P+n}$ and $g(z)=z^{P}$. Then, the component on sublattice B can be carried out by the Inverse Z-Transform, which reads as
	\begin{equation}
		\begin{aligned}\label{eq34}
			\braket{x_{2,B}}{\varphi_E}&=\frac{1}{2\pi i}\oint_{\abs{z}=1}z^{x_{2,B}-x_{1,A}-1}\Phi_B(z)dz\\ &=\frac{-\epsilon\braket{x_{1,A}}{\varphi_E}}{2\pi i}\oint_{\abs{z}=1}\frac{z^{x+P-1}}{f(z)}dz,
		\end{aligned}
	\end{equation}
	where we define $x=x_{2,B}-x_{1,A}$ as the relative distance. According to the {\it Fundamental Theorem of Algebra}, $f(z)$ can be factored as $f(z)=\prod_{\mu=1}^{P+W}(z-z_{\mu})\prod_{\nu=P+W+1}^{P+Q}(z-z_{\nu})$. Here we firstly assume that all $P+Q$ zeros of $f(z)$ are distinctive, i.e $z_{i}\ne z_{j}$ for $i\ne j$. The index $\mu$ ($\nu$) labels the zeros of $f(z)$ inside (outside) the unit circle in complex plane. The number of zeros inside (outside) the complex plane is given by $P+W$ ($Q-W$). This can be identified by recalling the definition of winding number $W=n_{\rm zeros}-n_{\rm poles}$, where $n_{\rm zeros}$ ($n_{\rm poles}$) is the number of zeros (poles) of $h(z)$ inside the unit circle. In our case, $n_{\rm zeros}$ equals to the number of zeros of $f(z)$ inside the unit circle and $n_{\rm poles}=P$, thus the number of zeros of $f(z)$ inside the unit circle $n_{\rm zeros}=W+n_{\rm poles}=W+P$ and the number of zeros of $f(z)$ outside the unit circle equals to $P+Q-n_{\rm zeros}$ that gives $Q-W$. Then one can use reside theorem to integrate out Eq.\,(\ref{eq34}). Let $x_c=1-P$, the normalized edge state can be expressed as
	\begin{align}
		\langle x|\varphi_E\rangle=\frac{1}{\mathcal{N}}\qty(\sum_{\mu=1}^{P+W}\Res[\frac{z_{\mu}^{x-x_c}}{f(z_{\mu})}]+\Res[\frac{0^{x-x_c}}{f(0)}]),
	\end{align}
	where $\mathcal{N}$ is the normalized constant. This formula can be further evaluated as $\langle x|\varphi_E\rangle=$
	\begin{equation}\label{eq38}
		\frac{1}{\mathcal{N}}\left\{
		\begin{aligned}
			&\sum_{\mu=1}^{P+W}\frac{e^{-(x-x_c)/\xi_{\mu}}}{\prod_{\alpha\ne\mu}(z_{\mu}-z_{\alpha})}\\
			&\sum_{\nu=P+W+1}^{P+Q}\!-\!\Res[\frac{z_{\nu}^{x-x_c}}{f(z_{\nu})}]\!=\!\sum_{\nu=P+W+1}^{P+Q}\frac{-e^{-(x-x_c)/\xi_{\nu}}}{\prod_{\beta\ne\nu}(z_{\nu}-z_{\beta})}
		\end{aligned}.
		\right.
	\end{equation}
	The upper (bottom) line holds for $x\ge x_c$ ($x<x_c$). From this formula, it's clear that the obtained interaction is the superposition of exponential decay components with the different decay length $\xi_{\mu,\nu}=-1/\ln(z_{\mu,\nu})$, and the number of components involving superposition is $P+W$ ($Q-W$) for $x\ge x_c$ ($x<x_c$). Now we are in the position to derive the PLE interaction from Eq.\,(\ref{eq38}). To this end, consider that $n$ out of $P+W$ zeros inside the unit circle coincide with $z_p$, namely a higher-order zero $z_p$ with order $n$ in $f(z)$. Then, the superposition of these $n$ components with the same decay length $\xi_{p}=-1/\ln(z_p)$ reduces to
	\begin{align}
		\lim_{z_{i_1},\cdots,z_{i_n}\to z_{p}}\sum_{\mu=1}^{n}\frac{e^{-(x-x_c)/\xi_{i_{\mu}}}}{\prod_{i_{\alpha}\ne i_{\mu}}(z_{i_{\mu}}\!\!-\! z_{i_{\alpha}})\prod_{\alpha'\ne\qty{i_1,\cdots,i_n}}(z_{i_{\mu}}\!\!-\! z_{\alpha'})}.
	\end{align}
	To carry out this limitation, let $z_{i_1}=z_p+h,\ z_{i_2}=z_p+2h,\cdots,\ z_{i_n}=z_p+nh$ and we have
	\begin{equation}
		\begin{aligned}\label{int_limit}
			&\lim_{h\to 0}\sum_{\mu=1}^{n}\frac{1}{\prod_{\alpha'\ne\qty{i_1,\cdots,i_n}}(z_{i_{\mu}}-z_{\alpha'})}\frac{(-1)^{n-\mu}(z_p+\mu h)^{x-x_c}}{(\mu-1)!(n-\mu)!h^{n-1}}\\&\propto\lim_{h\to 0}\frac{1}{h^{n-1}}\sum_{s=0}^{n-1}\sum_{\mu=1}^{n}\frac{(-1)^{n-\mu}z_p^{x-x_c}\mqty(x-x_c \\ s)h^s}{(\mu-1)!(n-\mu)! z^s_p}+\order{h}.
		\end{aligned}
	\end{equation}
	
	The summation over $\mu$ gives
	\begin{equation}
		\begin{aligned}
			\sum_{\mu=1}^{n}&\frac{(-1)^{n-\mu}z_p^{x-x_c}\mqty(x-x_c \\ s)h^s}{(\mu-1)!(n-\mu)!z_p^s} \\
			&=h^sz_p^{x-x_c-s}\mqty(x-x_c \\ s)S_2(s+1,n),
		\end{aligned}
	\end{equation}
	where $S_2(\bullet,\bullet)$ is the Stirling number of the second kind. The right-hand-side of this equation gives 0 for $s\ne n-1$ and $h^sz_{p}^{x-x_c-s}\mqty(x-x_c \\ s)$ for $s=n-1$. Therefore, the limitation in the right-hand-side of Eq.\,(\ref{int_limit}) is given by
	\begin{equation}
		\begin{aligned}\label{int_ple}
			\lim_{h\to 0}&\frac{h^{n-1}}{h^{n-1}}z_p^{x-x_c-n+1}\mqty(x-x_c \\ n-1)+\order{h}\\&=\frac{e^{-(x-x_c-n+1)/\xi_p}}{(n-1)!}\prod_{\mu=1}^{n-1}(x-x_c-\mu+1).
		\end{aligned}
	\end{equation}
	The PLE interaction manifests itself in the power-law factor $\prod_{\mu=1}^{n-1}(x-x_c-\mu+1)$, which originates mathematically from the existence of $n$-order zero of $h(z)$, and physically from the superposition of $n$ exponential decay components with the same decay length $\xi_p$. Note that, similar fashion can also apply to the interaction for $x<x_c$, e.g., one just needs to consider the higher-order zeros of $h(z)$ outside the unit circle, which leads to the PLE interaction with left chirality.
	
	Now generalizing the results to a generic lattice, whose distribution of zeros of the characteristic polynomial $h(z)$ is arbitrary. The d-d interaction mediated by such lattice can be rewritten as $\langle x|\varphi_E\rangle=$
	\begin{equation}\label{int_general}
		\left\{
		\begin{aligned}
			&\braket{x\ge x_c}{\varphi_E}=\frac{1}{\mathcal{N}}\sum_{\mu=1}\sum_{\alpha=0}^{n_{\mu}-1}r_{\mu\alpha}(x-x_c)^{\alpha}e^{-(x-x_c)/\xi_{\mu}}\\
			&\braket{x< x_c}{\varphi_E}=\frac{1}{\mathcal{N}}\sum_{\nu=1}\sum_{\beta=0}^{n_{\nu}-1}l_{\nu\beta}(x-x_c)^{\beta}e^{-(x-x_c)/\xi_{\nu}}
		\end{aligned}.
		\right.
	\end{equation}
	Here the indices $\mu$ and $\nu$ cover respectively all the distinctive zeros of $h(z)$ inside and outside the unit circle, associating with the orders $n_{\mu}$ and $n_{\mu}$. Thus $\sum_{\mu}n_{\mu}=P+W$, $\sum_{\nu}n_{\nu}=Q-W$, and $\xi_{\mu,\nu}=-1/\ln(z_{\mu,\nu})$ denote the decay length. The constants $r_{\mu\alpha}$ and $l_{\nu\beta}$ can be straightforwardly evaluated as
	\begin{subequations}\label{int_const}
		\begin{equation}
			\begin{aligned}
				r_{\mu\alpha}=\frac{\Theta(\Re\xi_{\mu})}{(n_{\mu}-1)!}&\sum_{i=\alpha-1}^{n_{\mu}-1}\mqty(n_{\mu}-1 \\ i)S_1(i,\alpha-1) \\ & \times e^{-i/\xi_{\mu}}f^{(n_{\mu}-i-1)}_{\mu}(z_{\mu}), \\
			\end{aligned}
		\end{equation}
		\begin{equation}
			\begin{aligned}
				l_{\nu\beta}=\frac{-\Theta(-\Re\xi_{\nu})}{(n_{\nu}-1)!}&\sum_{j=\beta-1}^{n_{\nu}-1}\mqty(n_{\nu}-1 \\ j)S_1(j,\beta-1) \\ & \times e^{j/\xi_{\nu}}f^{(n_{\nu}-j-1)}_{\nu}(z_{\nu}),
			\end{aligned}
		\end{equation}
	\end{subequations}
	where $f_{\mu}(z)=\prod_{\nu\ne\mu}(z-z_{\nu})^{-n_{\nu}}$, $S_1(\bullet,\bullet)$ is the Stirling numbers of the first kind, and $\Theta(x)$ is the Heaviside function. With the solution of edge state, we can evaluate the d-d interaction according to
	$J(x_{1,A},x_{2,B})\propto\braket{x}{\varphi_E}$.
	
	According to Eq.\,(\ref{int_ple}), the existence of $n$-order zero of $h(z)$, namely the superposition of $n$ exponential components with the same decay length will lead to the PLE interaction with maximum power-law exponent $n-1$. This gives the upper bound for the power-law exponents of PLE interaction in Eq.\,(\ref{int_general})
	\begin{align}
		\max_{\Re\xi_{\mu}>0}\qty{n_{\mu}}\le P+W,\ 
		\max_{\Re\xi_{\nu}<0}\qty{n_{\nu}}\le Q-W,
	\end{align}
	which recover Eq.\,(\ref{bound}) in main text.

	\subsection{Edge state without chirality}
	The bath Hamiltonian in the previous subsection possess chiral symmetry respected with $\sigma_z$, which results in the asymmetric distribution of the zero-energy edge state around the QE. Now, let's consider the other case of bath Hamiltonian $H_b(k)=d_y(k)\sigma_y+d_z(k)\sigma_z$ possessing chiral symmetry respected with $\sigma_x$, i.e., $\sigma_x^{-1}H_{b}(k)\sigma_x=-H_b(k)$. Since the chiral symmetry around $E=0$, the QE still acts equivalently as a hard-core potential $V=\epsilon\dyad{x_{1,A}}{x_{1,A}}$ with $\epsilon=\lambda^2/(E_q-\omega_0)=\infty$, which guarantees the zero energy solution $\qty[H_b+V]\ket{\varphi_E}=0\ket{\varphi_E}$. As a consequence of symmetry $\sigma_x$, the zero energy has equal components on each sublattice. This can be quickly proven as ($|\varphi_E\rangle=\sum_n(a_n\ b_n)^T\otimes|n\rangle$)
	\begin{align}
		H\sum_n\mqty(a_n\\b_n)\ket{n}=0=-H\sum_n\mqty(b_n\\a_n)\ket{n},
	\end{align}
	which results in $|a_n|=|b_n|$.
	
	To solve the eigenequation, we first perform an spin-rotation transformation $\mathcal{R}_{\boldsymbol{e_y}}(-\pi/2)=\oplus_{x=-L/2}^{L/2}R_{\boldsymbol{e_y}}(-\pi/2)$ with $R_{\boldsymbol{n}}(\theta)=\exp(i\theta \boldsymbol{\sigma}\cdot\boldsymbol{n}/2)$ that only acts on the internal degrees of bath, then we have
	\begin{align}\label{eq39}
		[H^{\rm rot}_b+V^{\rm rot}]\ket{\varphi^{\rm rot}_E}=0\ket{\varphi^{\rm rot}_E},
	\end{align}
	where the rotated potential
	\begin{align}
		V^{\rm rot}=\frac{\epsilon}{2}\mqty(\ket{x_{1,A}} & \ket{x_{1,B}})\mqty(1 & 1 \\ 1 & 1)\mqty(\bra{x_{1,A}} \\ \bra{x_{1,B}}),
	\end{align}
	and eigenstate
	\begin{align}
		\ket{\varphi^{\rm rot}_E}=\mathcal{R}^{\dagger}_{\boldsymbol{e_y}}\qty(-\frac{\pi}{2})\ket{\varphi_E}.
	\end{align}
	One can easily verify that the bulk Hamiltonian $H^{\rm rot}_b(k)=R_{\boldsymbol{e_y}}(-\pi/2)H_b(k)R^{\dagger}_{\boldsymbol{e_y}}(-\pi/2)$ has the form
	\begin{align}
		H^{\rm rot}_b(k)=-d_z(k)\sigma_x+d_y(k)\sigma_y\coloneqq\mqty(0 & h_{\rm rot}(k) \\ {h^*_{\rm rot}}(k) & 0),
	\end{align}
	which restores the chiral symmetry $\sigma_zH^{\rm rot}_b(k)\sigma^{-1}_z=-H^{\rm rot}_b(k)$.
	Two important results about the edge state $\ket{\varphi^{\rm rot}_E}$ can be already extracted. i) Our analysis for the spatial profile in last subsection also holds for $\ket{\varphi^{\rm rot}_E}$, since the rotated Hamiltonian $H^{\rm rot}_b$ has chiral symmetry respected with $\sigma_z$. ii) $\ket{\varphi^{\rm rot}_E}$ (namely $\ket{\varphi_E}$) has symmetric distribution around the atom since the parity symmetry of system is restored. To be more quantitative, Eq.\,(\ref{eq33}) is now rewritten as
	\begin{subequations}\label{eq43}
		\begin{equation}
			\begin{aligned}
				h_{\rm rot}(z)\Phi^{\rm rot}_B(z)+\frac{\epsilon}{2}\Big(&\braket{x_{1,A}}{\varphi^{\rm rot}_E}z^{-x_{1,A}} \\ &+\braket{x_{1,B}}{\varphi^{\rm rot}_E}z^{-x_{1,B}}\Big)=0,
			\end{aligned}
		\end{equation}
		\begin{equation}
			\begin{aligned}
				h_{\rm rot}^*(z^{-1})\Phi^{\rm rot}_A(z)+\frac{\epsilon}{2}\Big(&\braket{x_{1,A}}{\varphi^{\rm rot}_E}z^{-x_{1,A}} \\ &+\braket{x_{1,B}}{\varphi^{\rm rot}_E}z^{-x_{1,B}}\Big)=0.
			\end{aligned}
		\end{equation}
	\end{subequations}
	Thus, the rotated edge state $\ket{\varphi^{\rm rot}_E}$ on sublattice B has the same form as the expression we derived in the previous subsection. Moreover, $h_{\rm rot}^*(z^{-1})$ possesses inverse plus conjugate zeros and poles compared to $h_{\rm rot}(z)$, i.e., a pole at $z_i$ of $h_{\rm rot}(z)$ corresponds to a pole at $1/z^*_i$ of $h_{\rm rot}^*(z^{-1})$ and vice versa. The distribution for zeros and poles between $h_{\rm rot}(z)$ and $h_{\rm rot}^*(z^{-1})$ eventually leads to the inverse plus conjugate spatial profile between $\braket{x_{2,A}}{\varphi^{\rm rot}_E}$ and $\braket{x_{2,B}}{\varphi^{\rm rot}_E}$, i.e.,
	\begin{align}
		\braket{x_A}{\varphi^{\rm rot}_E}=\braket{\varphi^{\rm rot}_E}{x_B},\ x_A=-x_B,
	\end{align}
	where the relative distance $x_{A}=x_{2,A}-x_{1,A}$ and $x_{B}=x_{2,B}-x_{1,B}$. The final step to obtain the edge state is a inverse rotation to $\braket{x_{A/B}}{\varphi^{\rm rot}_E}$
	\begin{align}
		\ket{\varphi_E}=\oplus_{x}R_{y}\qty(-\frac{\pi}{2})\mqty(\braket{x_A}{\varphi^{\rm rot}_E}\ket{x_A} \\ \braket{x_B}{\varphi^{\rm rot}_E}\ket{x_B}),
	\end{align}
	which gives the edge state without chirality
	\begin{subequations}
		\begin{align}
			\abs{\braket{x_A}{\varphi_E}}=\frac{1}{\sqrt{2}}\abs{\braket{\varphi^{\rm rot}_E}{-x_A}-\braket{x_A}{\varphi^{\rm rot}_E}}=\abs{\braket{-x_A}{\varphi_E}},
		\end{align}
		\begin{align}
			\abs{\braket{x_B}{\varphi_E}}=\frac{1}{\sqrt{2}}\abs{\braket{\varphi^{\rm rot}_E}{-x_B}+\braket{x_B}{\varphi^{\rm rot}_E}}=\abs{\braket{-x_B}{\varphi_E}}.
		\end{align}
	\end{subequations}
	The obtained edge states are just the linear combination of the edge states we obtained in previous subsection, thus the corresponding power-law exponent can only either be invariant or decreasing. The upper bound for right-side component $\langle x_{A/B}\ge0|\varphi_E\rangle$ is determined by $\max\qty{P_{\rm rot}+W,Q_{\rm rot}-W}$ where $P_{\rm rot}$ and $Q_{\rm rot}$  are the hopping parameters associated with $H^{\rm rot}_b$, and $W$ is the winding number corresponding to $H^{\rm rot}_b$. The same conclusion holds for left-side component due to the parity symmetry of edge states.
	The same conclusion holds for left-side component due to the parity symmetry of edge states.
	\begin{figure}
		\centering
		\includegraphics[width=8.4cm]{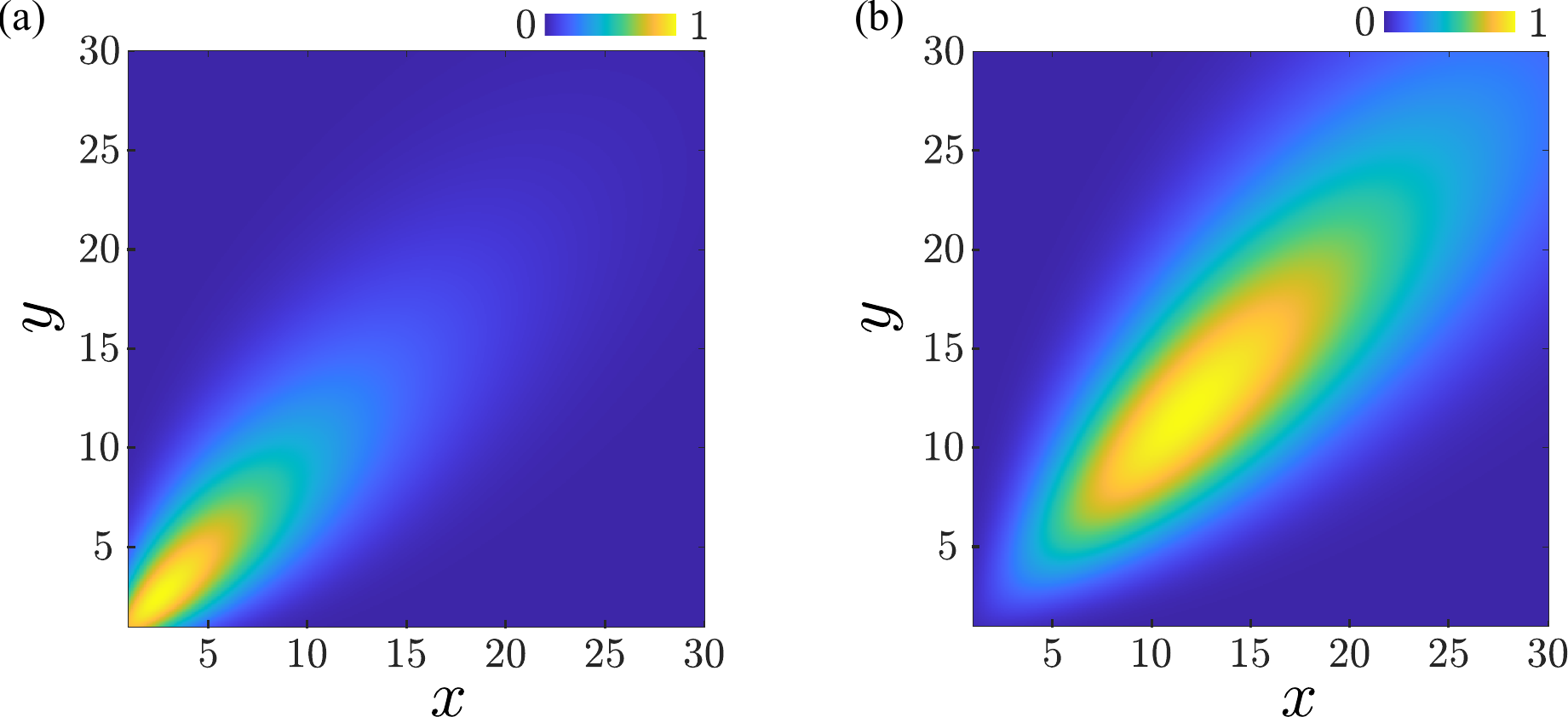}
		\caption{Interaction obtained from Eq.\,(\ref{int2d_1}) and Eq.\,(\ref{int2d_2}) are shown in (a) and (b), respectively. The interaction strengths were rescaled to its maximum values. The factorial function has been replaced by the gamma function to obtain the continuous spatial profile.}\label{fig3}
	\end{figure}
	
	\section{PLE interaction with the higher power-law exponent}\label{suppB}
	Our analytical derivation in the main text claims that the largest power-law exponent of interaction is upper bounded by $P+W-1$ (here we consider the completely chiral case which implies $Q-W=0$ for simplicity). Therefore, one firstly need to increase $P+W$ to obtain a PLE interaction with a higher power-law exponent. For example, one can add the long (but finite)-range hopping to increase $P$. After that, the bath needs to be fine-tuning to generate the high-order zeros of $h(z)$ in order to obtain a higher power-law exponent for a given bound.
	
\begin{figure}
		\centering
		\includegraphics[width=8.2cm]{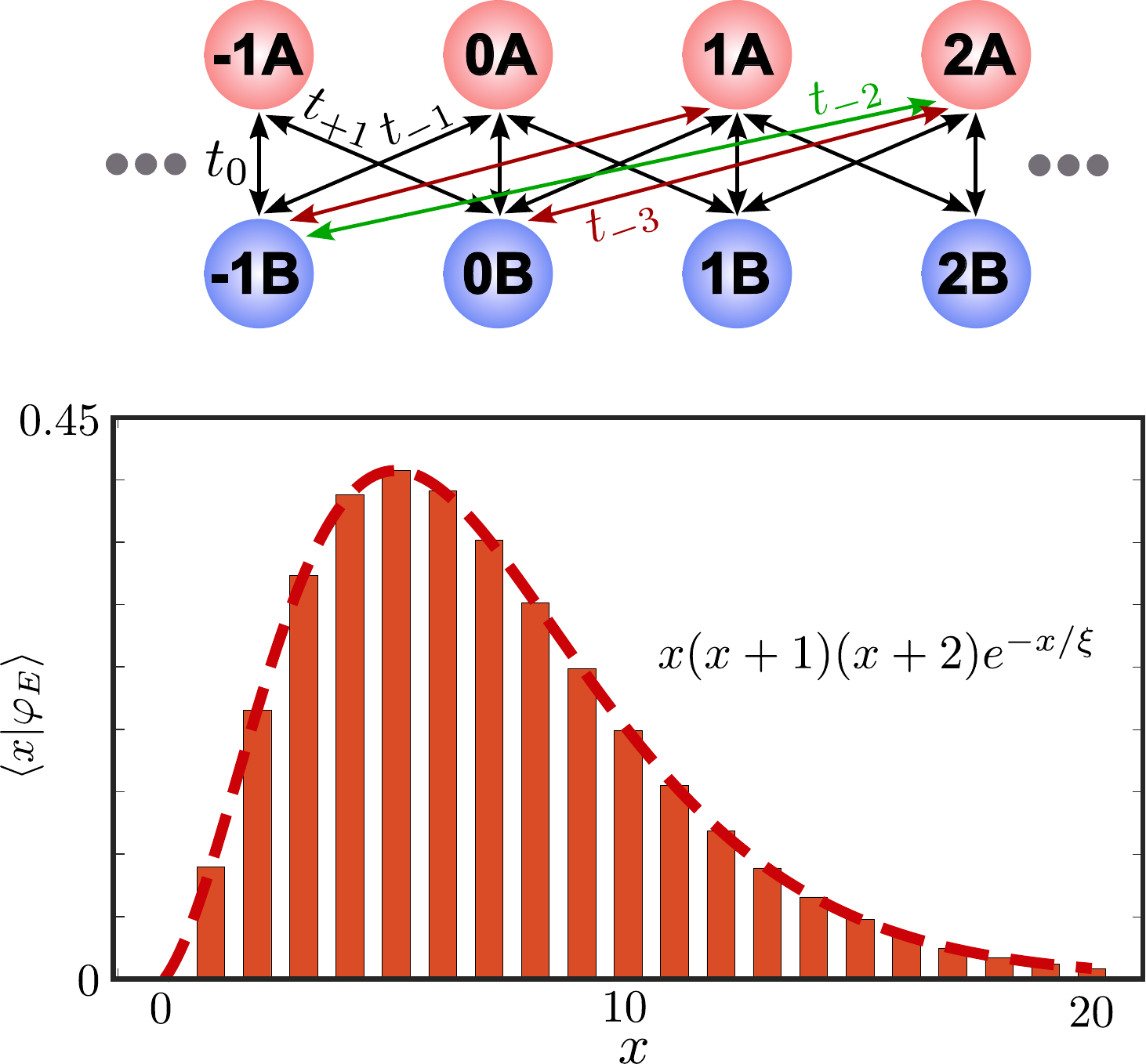}
		\caption{Bottom plot shows the chiral PLE edge state corresponded to bath Hamiltonian with $h(k)=\qty(e^{ik}-3/5)^4/e^{3ik}$, which corresponds to the chiral PLE d-d interaction with $\xi=-1/\ln(0.6)$. The histogram and dashed line is drawn from the numerical and analytical solution, respectively. Upper plot shows the configuration with $t_{-3}/t_0=-27/500,\ t_{-2}/t_0=-9/25,\ t_{-1}/t_0=-9/10,\ t_{+1}/t_0=-5/12$.}\label{fig5}
	\end{figure}

	Here we illustrate the PLE interaction with maximal power-law exponent of $3$. This requires the next-next-nearest neighbor hopping between cells of bath [see Fig.\,\ref{fig5}]. As observed in Fig.\,\ref{fig5}, the PLE decay edge state is obtained by tuning bath parameters to $h(k)=\qty(e^{ik}-3/5)^4/e^{3ik}$, as shown in Fig.\,\ref{fig5}(a).
	
	\section{Power-law enhanced interaction in higher dimensional lattice}\label{suppC}
	In this section, we demonstrate the implement of power-law enhanced interaction in 2D square lattice. Our conclusion can also be applied to higher-dimensional lattice bath. We do not refer to ``PLE" interaction for reasons will be given latter. For theoretical simplification, the 2D square lattice under consideration respect chiral symmetry $\sigma_z$, i.e., $\sigma_zH(k_x,k_y)\sigma_z=-H(k_x,k_y)$, such that the VDS mechanism is valid and the equations for state components on different sublattice are decoupled. The full properties of the lattice are encoded in $h(k_x,k_y)$. We assume the QE couples to lattice in sublattice A of position $(0,0)$. It is known that a gappless lattice in higher dimension can also produce coherent d-d interaction, due to the vanishing density of states\,\cite{gapless1,gapless2}. However, we here consider the gapped lattice, which means that $h(z_1,z_2)\ne0$ for any point $z_1,z_2$ on the 2D circle surface. Taking similar fashion in section II, we have
	\begin{equation}
		\begin{aligned}\label{int2d}
			J(\boldsymbol{r}_B)&\propto\langle \boldsymbol{r}_B|\varphi_E\rangle \\ &\propto\frac{1}{(2\pi i)^2}\oint_{\abs{z_1}=1}\oint_{\abs{z_2}=1}\Phi_B(z_1,z_2)z^{x-1}_1z^{y-1}_2dz_1dz_2
		\end{aligned}
	\end{equation}
	
	and $\Phi_A(z_1,z_2)=0$, where $\boldsymbol{r}_{A/B}=(x,y)$ denotes the position vector in sublattice $A/B$. The integral contours can be chosen to be the individual unit circle since we consider a gapped lattice. One can easily show that the generic form of $\Phi_B(z_1,z_2)$ is
	
	\begin{align}
		\Phi_B(z_1,z_2)=\frac{1}{h(z_1,z_2)}=\frac{1}{\sum_{ij}t_{i,j}z^{i}_1z^{j}_2},
	\end{align}
	where $h(z_1,z_2)\equiv h(k_x\to-i\ln z_1,k_y\to-i\ln z_2)$. The expression for $h(z_1,z_2)$ can be obtained as i) a $p_x$-site ($q_x$-site) hopping from sublattice A to B towards left (right) direction contributes a $z^{-p_x}_1$ ($z^{q_x}_1$); ii) a $p_y$-site ($q_y$-site) hopping from sublattice A to B towards up (down) direction contributes a $z^{-p_y}_2$ ($z^{q_y}_2$). Thereby, the coefficient $t_{ij}$ denotes the hopping strength of lattice.
	
	Unlike the 1D lattice we consider in main text, in which the form of dressed bound state can be explicitly solved for arbitrary configuration of lattice owing to the {\it Fundamental Theorem of Algebra}, as long as the hopping range is finite. In 2D geometry, this theorem fails for a two variables characteristic polynomial $h(z_1,z_2)$, thus Eq.\,(\ref{int2d}) can barely be evaluated even for simple lattice configuration. Nonetheless, the analog of PLE interaction in 1D lattice, namely the power-law enhanced interaction, can be obtained as
	\begin{equation}
		\begin{aligned}
			J'(\boldsymbol{r}_B)=x^{\alpha}y^{\beta}&J(\boldsymbol{r}_B)\propto\frac{1}{(2\pi i)^2}\oint_{\abs{z_1}=1}\oint_{\abs{z_2}=1}z^{x-1}_1z^{y-1}_2\\ &\times \mathcal{D}^{\alpha}(z_1)\mathcal{D}^{\beta}(z_2)\Phi_B(z_1,z_2)dz_1dz_2,
		\end{aligned}
	\end{equation}
	where we introduce the operator $\mathcal{D}(z)=z\partial/\partial z$. In other word, the d-d interaction mediated by the new lattice with the characteristic function $h'(z_1,z_2)=(\mathcal{D}^{\alpha}(z_1)\mathcal{D}^{\alpha}(z_2)h^{-1}(z_1,z_2))^{-1}$ has power-law enhancement $x^{\alpha}y^{\beta}$ compared with the d-d interaction in the original lattice with $h(z_1,z_2)$. Clearly, one can regard the PLE interaction in 1D as a special case of power-law enhanced interaction, where the original exponential decay interaction $J(x)$ are enhanced by the power-law factor, thus leads to $J'(x)\propto x^{\alpha}\exp(-x/\xi)$. Despite the power-law enhanced behavior similar to PLE interaction in 1D, there are no corresponding upper bounds for power-law exponents $\alpha,\beta$ in 2D lattice. This is again, due to the failure of {\it Fundamental Theorem of Algebra} in 2D geometry.

	We present a solvable model to demonstrate the power-law enhanced interaction in 2D. To this end, we consider an original lattice with $h(k_x,k_y)=t_{0,0}+t_{-1,0}\exp(-ik_x)+t_{0,-1}\exp(-ik_y)$. When hopping strengths satisfy $|t_{-1,0}|+|t_{0,-1}|<|t_{0,0}|$, the resulted d-d interaction can be obtained as
	\begin{align}\label{int2d_1}
		J(\boldsymbol{r}_B)\propto e^{-x/\xi_x}e^{-y/\xi_y}F(x,y),\ x,y\ge 0,
	\end{align}
	where the interaction length $\xi_x=-\ln(t_{-1,0}/t_{0,0})$ and $\xi_y=-\ln(t_{0,-1}/t_{0,0})$, $F(x,y)=(x+y)!/x!y!$. The power-law enhanced interaction with power-law exponent 1 in both $x$ and $y$ direction
	\begin{align}\label{int2d_2}
		J'(\boldsymbol{r}_B)\propto xye^{-x/\xi_x}e^{-y/\xi_y}F(x,y),\ x,y\ge 0
	\end{align}
	can be mediated by the lattice with bulk Hamiltonian $h'(k_x,k_y)=(t_{0,0}+t_{-1,0}\exp(-ik_x)+t_{0,-1}\exp(-ik_y))^3\exp(ikx\!+\!iky)$. As shown in Fig.\,\ref{fig3}, the spatial profiles of $J'(\boldsymbol{r}_B)$ exhibits non-monotonic behavior in this case, just as the PLE interaction in 1D lattice.
	
	\section{FM phase and its boundary}\label{suppD}
	We analytically calculate the FM phase boundary via a spin-wave analysis. For $J_z\ll0$, one should expect all spins are polarized along the $+z$ direction. Thus, we treat $|\uparrow\uparrow\cdots\uparrow\rangle$ as the vacuum state with no excitations and apply the Holstein-Primakoff transformation $S^z_j=1/2-a^{\dagger}_ja_j$, $S^{+}_j=a_j$, $S^-=a^{\dagger}_j$, where $[a_i,a^{\dagger}_j]=\delta_{ij}$. 
	
	In the weak excitation limit, $\langle a^{\dagger}_ja_j\rangle\ll1$, the XXZ Hamiltonian under PLE interaction becomes
	\begin{equation}
		\begin{aligned}
			H_{\rm XXZ}\approx\frac{1}{2}\sum_{m>n}&(m-n)e^{-(m-n)/\xi}\Big(a^{\dagger}_ma_n+a^{\dagger}_na_m \\ &-J_z(a^{\dagger}_ma_m+a^{\dagger}_na_n)\Big).
		\end{aligned}
	\end{equation}
	
	In the thermodynamic limit $N\to\infty$, one can diagonalize this Hamiltonian to $H_{\rm PLE}=\int dk\omega_kc^{\dagger}_kc_k$ for $k\in (-\pi,\pi]$ with the following dispersion relation
	\begin{align}\label{wk1}
		\omega_k\!=\!\frac{-J_ze^{-1/\xi}}{(-1+e^{-1/\xi})^2}\!+\!\frac{e^{-1/\xi}\qty((1+e^{-2/\xi})\cos(k)-2e^{-1/\xi})}{(1+e^{-2/\xi}-2e^{-1/\xi}\cos(k))^2}.
	\end{align}
	The $\omega_{\rm min}\equiv \min\omega(k)=0$ condition sets the phase boundary. For dispersion Eq.\,(\ref{wk1}), $\omega_{\rm min}=\omega(k=\pi)$ for $\xi\le -1/\log(2-\sqrt{3})$ and $\omega_{\rm min}=\omega(k=\pm k_0)$ for $\xi> -1/\log(2-\sqrt{3})$ where
	\begin{align}
		k_0=\arctan(\frac{\sqrt{-(e^{-4/\xi}-14e^{-2/\xi}+1)(e^{-2/\xi}-1)^2}}{-e^{-4/\xi}+6e^{-2/\xi}-1}).
	\end{align}
	Then, the boundary can be obtained by solving $\omega_{\rm min}=0$, which gives
	\begin{align}
		J_z=\frac{(e^{-1/\xi}-1)^2\qty((1+e^{-2/\xi})\cos(\Re(k_0))-2e^{-1/\xi})}{(1+e^{-2/\xi}-2e^{-1/\xi}\cos(\Re(k_0)))^2}.
	\end{align}
	
	For exponential decay and power-law decay interacting systems, similar fashion applied and we can obtain the corresponding dispersion relations as
	\begin{align}\label{wk2}
		\omega_k=\frac{-J_ze^{-1/\xi}}{-1+e^{-1/\xi}}-\frac{e^{-1/\xi}\qty(1+e^{2ik}-2e^{ik-1/\xi})}{2(e^{ik}-e^{-1/\xi})(-1+e^{ik-1/\xi})}
	\end{align}
	and
	\begin{align}\label{wk3}
		\omega_k=-J_z\sum_{r=1}^{\infty}\frac{1}{r^{\xi}}+\sum_{r=1}^{\infty}\frac{\cos(kr)}{r^{\xi}},
	\end{align}
	respectively. The minima of $\omega_k$ for both exponential decay and power-law decay interactions appear at $k=\pi$. Therefore, the boundary can be solved as
	\begin{align}
		J_z=\frac{e^{-1/\xi}-1}{e^{-1/\xi}+1}
	\end{align}
	and
	\begin{align}\label{pl_b}
		J_z={\rm Li}_{\xi}(-1)/\zeta(\xi),\ \xi>1,
	\end{align}
	respectively. Here ${\rm Li}_{s}(z)$ the polylogarithm function and $\zeta(z)$ the Riemann zeta function. Note that, the obtained boundary for power-law decay interaction is $J_z=0$ for $\xi\le1$, since the FM state’s energy is super extensive\,\cite{GZX}.


\begin{thebibliography}{56}
\bibitem [{def()}]{def}%
	\BibitemOpen
	\href@noop {} {}\bibinfo {note} {The PLE interaction implies
		$h(k)=(\exp(ik)-\exp(-1/\xi))^2/\exp(ik)$. The band gap $\Delta_g$ equals to
		$2h(0)$. After some calculations one can obtain the expression in main
		text.}\BibitemShut {Stop}%
\end{thebibliography}
\end{document}